\documentclass[pre,aps,twocolumn,showpacs,floatfix]{revtex4-1}

\usepackage{dcolumn}
\usepackage{amsmath}
\usepackage{amssymb}
\usepackage{bm}
\usepackage{graphicx,subfigure}
\usepackage{epsf,epsfig}
\usepackage{hyperref}
\usepackage{color}
\usepackage{enumerate}

\begin{document}
\title{Sponge-like rigid structures in frictional granular packings}
\author{Kuang Liu$^1$, Jonathan E. Kollmer$^{2,3}$, Karen E. Daniels$^3$, J. M. Schwarz$^{1,4}$, Silke Henkes$^5$}
\affiliation{$^1$ Physics Department, Syracuse University, Syracuse, NY USA,
$^2$ Department of Physics, Universit\"at Duisburg-Essen, Germany,
$^3$ Department of Physics, North Carolina State University, Rayleigh, NC USA,
$^4$ Indian Creek Farm, Ithaca, NY USA,
$^5$ School of Mathematics, University of Bristol, Bristol, England, United Kingdom}
\date{\today}
\begin{abstract}
We show how rigidity emerges in experiments of sheared frictional granular materials by using generalizations of two methods for identifying rigid structures. Both approaches, the force-based dynamical matrix and the topology-based rigidity percolation, agree with each other and identify similar rigid structures.  As the system becomes jammed, at a contact number of $z=2.4\pm 0.1$, a rigid backbone interspersed with floppy, particle-filled holes of a broad range of sizes emerges, creating a sponge-like morphology. We also find that the pressure within rigid structures always exceeds the pressure outside the rigid structures, i.e. that the backbone is load-bearing. These findings shows that it is necessary to go beyond mean-field theory to capture the physics of frictional jamming and also suggests that mechanical stability arises through arch structures and hinges at the mesoscale.
\end{abstract}                                                                                   
\maketitle

Rigidity is the ability of a system to resist imposed perturbations; for disordered materials, their detailed internal structure determines rigidity~\cite{Alexander1998,Lubensky2015}. The Maxwell counting criterion~\cite{Maxwell1864}, first developed for building girder frameworks in $19^{\text{th}}$ century railway bridges, has long been used to compute the stability of system by comparing the number of constraints to the number of degrees of freedom~\cite{Calladine}.
This simple, effectively mean-field, criterion also correctly predicts the onset of positive bulk and shear moduli in frictionless jamming of spherical particles (e.g. foams or emulsions)~\cite{OHern2003,van2009jamming,DagoisBohy2012,Goodrich2014}. However, when friction is introduced, as required for modeling granular materials, the counting argument no longer works, even with modifications \cite{silbert2002geometry, shundyak2007force, bi2011jamming}. In particular, systems which acquire rigidity under shear do so at lower packing fractions than those loaded isotropically, via the appearance of anisotropic, load-bearing force chains, in a phenomenon known as shear-jamming~\cite{bi2011jamming}. These findings highlight the importance of local structure, and raise the question of the suitability of a mean-field criterion. The common practice of counting Maxwell constraints only after removing the \emph{rattlers} \cite{OHern2003,majmudar2007jamming} (particles with few contacts) has further obscured this issue.

A first approach to local rigidity, linear response theory, uses the detailed local geometry and forces to compute the {\it dynamical matrix} (or Hessian) of the system~\cite{silbert2005vibrations}. A rigid packing will have no system-spanning zero-modes in the dynamical matrix, except for global translations and rotations; conversely their presence indicates a lack of rigidity. In frictionless systems, this method agrees with the result of Maxwell constraint counting, after removing rattlers~\cite{silbert2005vibrations}. 
In frictional systems, the same comparison was made using a dynamical matrix extended to include friction~\cite{somfai2007critical,henkes2010}. In simulations of frictional packings equilibrated at constant pressure, the results from this extended dynamical matrix match a generalised form of the constraint counting argument, creating a frictional jamming transition along a generalised isostaticity line~\cite{shundyak2007force, henkes2010}. Other modified isostatic conditions have been proposed for frictional systems \cite{Liu2017,Ikeda2020}, but none have yet been experimentally tested. 

A second approach to quantifying local rigidity focuses on the spatial patterns of rigid clusters, which are sets of connected bonds that are mutually rigid with respect to one another~\cite{Feng1984}. In this \emph{rigidity percolation} framework, the rigidity transition corresponds to the emergence of a spanning rigid cluster in the contact network. In 2D, the \emph{pebble game}~\cite{jacobs1997algorithm} uses Laman's theorem to construct a generic algorithm for decomposing a network into rigid clusters and floppy regions. This theorem depends only on the network topology and does not require information about forces and contact geometry. Analysis of 2D systems show that frictionless packings exhibit a discontinuous rigidity transition~\cite{ellenbroek2015rigidity}, while generic central-force networks exhibit a continuous transition~\cite{Jacobs95,Jacobs96}. Rigidity percolation has also provided insights into the structure of colloidal gels with attractive interactions \cite{Koeze2029sticky,zhang2019correlated,Lester2}.
Recent work \cite{henkes2016} has extended the pebble game to frictional packings and showed that networks derived from slowly-sheared frictional simulations generate rigid cluster structures consistent with a continuous transition. Using a simplified network model, we have additionally established the transition as continuous, but with exponents that differ from standard (central-force) rigidity percolation \cite{liu2019}. Simulations on shear-jammed states further indicate that the onset of shear jamming corresponds to the percolation of overconstrained regions with a broad range of sizes~\cite{Vinutha2019}. Experimental tests of rigidity percolation for frictional systems are absent.

\begin{figure*}
	\centering
	\includegraphics[width=1.0\textwidth]{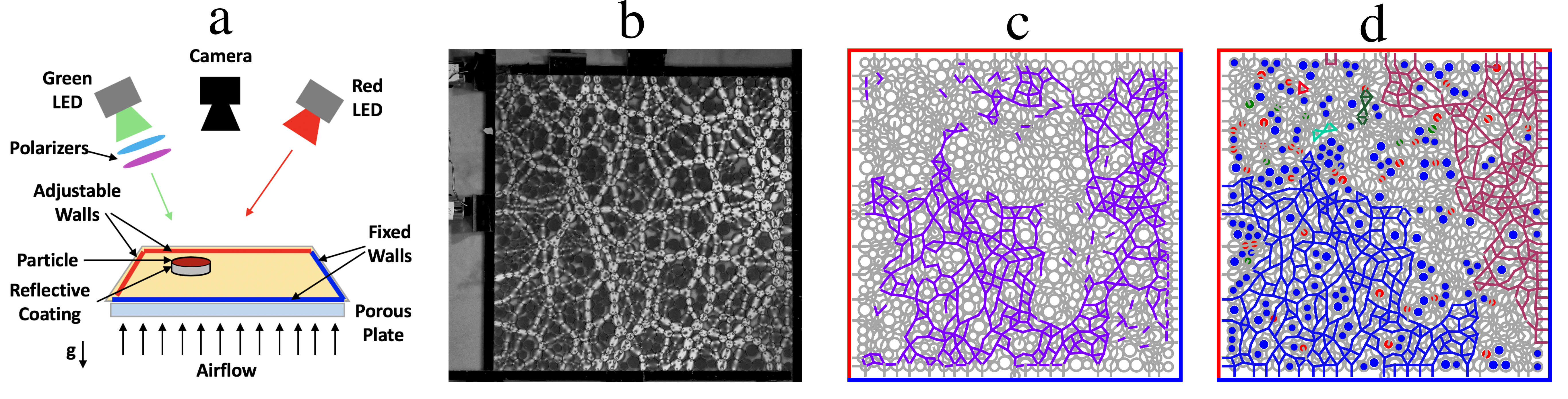}
	\caption{(a) Schematic of the experimental setup with fixed walls (blue) and moving walls (red). 
		(b) Sample image, showing just the polarized channel (photoelastic respose).
		(c) Rigid region decomposition of sample (b), computed using the dynamical matrix; the rigid region is purple, and floppy bonds are grey.
		(d) Rigid cluster decomposition of sample (b), computed using the pebble game; it contains two large rigid clusters (blue and red bonds), some smaller rigid clusters (other colors), and regions of floppy bonds (grey). Blue particles are rattlers with zero constraints.}
	\label{fig:schematic}
\end{figure*}

In this Letter, we apply both the dynamical matrix and the pebble game to data from experiments on 2D {\it frictional} granular packings. We measure particle positions and forces within a monolayer of quasi-statically sheared grains floating on a gentle cushion of air \cite{puckett2013,bililign_protocol_2019}, with interparticle forces obtained using photoelasticity \cite{daniels_photoelastic_2017, abed_zadeh_enlightening_2019}.
We find that both the frictional dynamical matrix and the frictional pebble game agree with each other, providing nearly identical decompositions of the packings into rigid and floppy regions, and that there is a strong correlation between local pressure and local rigidity. The transition in our finite-sized system occurs at $z_c = 2.4 \pm 0.1$, well below the mean field value $z_c=3$. We discover that the rigid structures are sponge-like, i.e. containing a broad range of floppy hole sizes, particularly near the transition, which is again a signature of a continuous transition inconsistent with mean-field rigidity.

{\it Experiments:} We perform experiments on a monolayer of $N=826$  photoelastic bidisperse disks (Fig.~\ref{fig:schematic}a). The two particle radii are $R_1 = 5.5$~mm and $R_2 = 7.7$~mm (with $R_2/R_1 = 1.4)$),  and the particles are initially confined to an area of approximately $L = 0.5 \times 0.5$~m$^2$. Two of the confining walls are controlled by stepper motors; to impose simple shear, one wall moves in while the other moves out in a series of quasi-static steps of size $\Delta x = 1.5$~mm, with $\Delta y$ adjusted to maintain constant area $A$. After $n$ steps, each resulting in a shear strain $\epsilon = \frac{\Delta x}{L} \approx 0.003$, the shear is reversed back to the initial state. The number of steps is not fixed, but ranges from $n=8$ (ending at a total stress threshold)  to $n=13$ (pre-defined maximum). The floor of the shear cell is a porous frit through which air flows to allow the particles to float on a gentle air cushion, creating a system without basal friction; this apparatus is largely the same as the one described in \cite{puckett2013,bililign_protocol_2019}. Therefore, the external load from the two walls is the only significant external stress. The complete dataset consists of $24$ cyclic runs, with each run starting from randomized particle positions and an initial barely-jammed volume. The packing fraction for each of the 24 runs is in the range $0.746< \phi < 0.760 \pm 0.006$. During each cycle, contacts are created through shear during  the first half of the cycle (dubbed `shear'), and partially released during the second half of the cycle (dubbed `unshear'); due to shear-jamming \cite{bi2011jamming,zheng_shear_2014}, the system does not return to its initial state after a complete cycle. Datasets where we could not track all particles where discarded. A total of 353 images are used in the analysis below. 
Since the particles are made of a birefringent material (Vishay PhotoStress PSM-4), we are able to use photoelasticity \cite{daniels_photoelastic_2017,abed_zadeh_enlightening_2019} to measure the vector contact forces on all particles; a sample image is shown in Fig.~\ref{fig:schematic}b. The red channel (not shown) uses unpolarized light and measures particle positions, and the green channel (shown) uses circularly polarized light to measure the photoelastic signal. From the later, we  determine the normal and tangential contact forces ($f_n, f_t$) on each particle using our open-source algorithms \cite{daniels_photoelastic_2017,PEGS}. From measurements of the normal $f_n$ and tangential $f_t$ contact forces, we estimate a friction coefficient of $\mu = 0.3$ (see \cite{SM}). The Coulomb threshold for the mobilisation $|f_t|/\mu f_n$ determines whether a contact is sliding ($\geqslant 1$) or frictional ($<1$); its distribution has so far only been analysed in simulations \cite{shundyak2007force}. The rigidity calculations, described below, depend sensitively on the correct determination of whether two particles are in contact. The SM \cite{SM} provides information on how we determine the optimal parameters. In all cases, we find that values of the mean coordination number are known to within $\pm 0.1$.

\begin{figure*}
	\centering
	\includegraphics[width=1.0\textwidth]{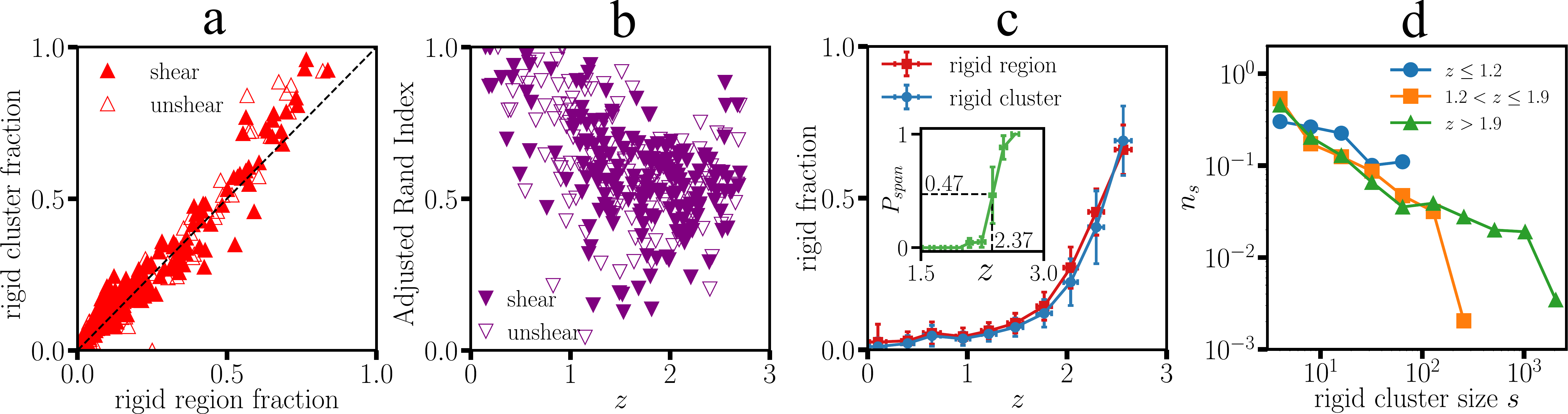}
	\caption{Correlations between rigid clusters and regions, calculated on the 353-image dataset. (a) Correlation between the rigid cluster fraction and the rigid region fraction. (b) Adjusted Rand index (ARI) between the rigid cluster decomposition and the rigid region decomposition. (c) Fractions of rigid clusters and rigid regions as function of average coordination number, $z$. Inset: Probability of a spanning rigid cluster. (d) Histogram of  cluster size $s$, taken for three different ranges of $z$. }
	\label{fig:fraction}
\end{figure*}

{\it Rigidity computations:} We first compute the vibrational modes of the system, starting by expanding the equations of motion about mechanical equilibrium
\begin{equation}
\delta\ddot{r}^i_\alpha =-D^{ij}_{\alpha\beta}\delta r^j_{\beta}+\text{dissipation}(\delta \dot{r}) + O(\delta r^2),
\end{equation} 
where $ D^{ij}_{\alpha,\beta}  =  \frac{1}{\sqrt{m_{i,\alpha} m_{j,\beta}}}\frac{\partial^2 V_{ij}}{\partial r_{i,\alpha} r_{j,\beta}}$ is the dynamical matrix of the system, the indices $(i,j)$ label all disks, $(\alpha,\beta)$ label the two spatial $x,y$ components and the angular component $R\theta$,  and $m$ denotes the particle mass.
While frictional interactions are not conservative, one can nevertheless derive an effective potential in linear response (see \cite{SM}). We arrive at $V_{\text{eff}}^f =\frac{1}{2} K_t \delta t^2$ for a contact with stiffness $K_t$ during  tangential contact loading, where $\delta t$ is the tangential displacement at the contact point. 
For a sliding contact at the Coulomb threshold, we approximate that the shear dynamics does not reverse the sliding direction. We have verified this assumption in sheared simulations~\cite{unpublished} and do not include the shear-reversal step in our analysis here. We obtain $V_{\text{eff}}^f =\pm \mu f_n \delta t$, where $f_n$ is the value of the normal force at equilibrium.  
Then the effective potential  becomes \cite{somfai2007critical,henkes2010,SM}
\begin{equation}
 V_{ij} = \frac{1}{2} \left[ K_n (\delta \mathbf{r} \cdot \hat{\mathbf{n}})^2 - f_n/|\mathbf{r}_{ij}| \left( \delta \mathbf{r} \cdot \hat{\mathbf{t}}\right)^2 + V_{\text{eff}}^f \right],  
 \label{eq:potential}
\end{equation}
with normal elastic stiffness $K_n$, and where the third term arises only for friction. 
To construct the dynamical matrix for our experimental data, we use measured masses for $m$, and estimate $K_n$ from the elastic modulus of the material. To ensure that tangential and normal interactions contribute at the same order, we set $K_t=K_n$. Using the particle positions and interparticle forces, we then construct the dynamical matrix and compute its normalized eigenmodes. The zero eigenvalue modes parametrise the floppy motions, and we determine translational and rotational relative displacements at contacts between disk pairs.  We then compute the mean square displacement over floppy modes at individual bonds and mark all bonds with a displacement below (above) a threshold value $2\cdot 10^{-5}$ as rigid (floppy); there is mild threshold dependence \cite{SM}.
In the transition region, we obtain sets of contiguous rigid bonds that form {\it rigid regions}, shown in Fig.~\ref{fig:schematic}c. 

Our second method of measuring rigidity is to decompose the system into rigid clusters using the frictional pebble game. To do so, we extend the central force $(k\!=\!2,l\!=\!3)$ pebble game applied to a contact network to a ($k\!=\!3,l\!=\!3$) pebble game in order to incorporate the additional rotational degree of freedom made relevant by the friction between disks. Moreover, each contact below the Coulomb threshold contributes two constraints (one normal and one tangential), while each contact at the threshold (freely sliding) only contributes a normal constraint \cite{henkes2016,liu2019}. To this constraint network, we add an appropriate number of constraint bonds between the four boundaries in the experiment and all contacting particles (see \cite{SM}). A sample decomposition is shown in  Fig.~\ref{fig:schematic}d.

{\it Results:} Using the particle positions and inter-particle forces obtained from experiments, we apply the dynamical matrix method and the frictional pebble game to determine \emph{rigid regions} and \emph{rigid clusters}, respectively. Fig.~\ref{fig:schematic}c-d, performed on an image near the onset of jamming, illustrates that the identified rigid clusters/regions are closely correlated. This correspondence remains true for our full dataset: Fig.~\ref{fig:fraction}a is a scatter plot of the measured rigid cluster fraction against the rigid region fraction. All data points are clustered around the diagonal, with no difference between the shear and unshear directions. We find that the pebble game detects a slightly higher rigid fraction at high $z$, possibly due to boundary effects. This system-scale correspondence carries over to the contact level (Fig.~\ref{fig:fraction}b), where we compute the adjusted Rand index (ARI) \cite{Rand1971,Hubert1985,SM} to measure the bond-scale similarity of the detected clusters/regions.
We find $\mathrm{ARI} > 0$ (correlation is present), with an average of $0.6$ indicating strong positive correlation and some differences again apparent at higher $z$. This robust \cite{SM} high degree of correspondence is significant since the rigid cluster method requires only information about the contact graph (it is simply a \emph{topological} measure), in contrast to the explicit displacement computation in the dynamical matrix, which contains the full spatial information. The correspondence is not exact, and there are specific (known, but rare) configurations where the two approaches give different results \cite{Lester2018}.
In  Fig.~\ref{fig:fraction}c, we observe that the rigid cluster/region fractions both indicate a continuous rigidity transition, and agree with each other within error bars. Using the probability of a spanning rigid cluster (inset), we measure the transition point of $z_c=2.4\pm 0.1$. Fig.~\ref{fig:fraction}d shows that the rigid cluster size distribution broadens with increasing $z$. While our data is limited by finite system size and finite statistics, our distributions do not have a gap, and strongly resemble the results found in simulations of frictional disks~\cite{henkes2016}. These findings are consistent with a continuous rigidity percolation transition at a value of $z_c<3$, the mean-field Maxwell criterion with friction, and also with the mechanism of shear jamming. Note that in all of our analyses, we do not remove rattlers, as they are an integral part of the coexisting floppy and rigid regions and can become part of the packing at some point during the shear.

\begin{figure}[t]
	\centering
	\includegraphics[width=0.9\columnwidth]{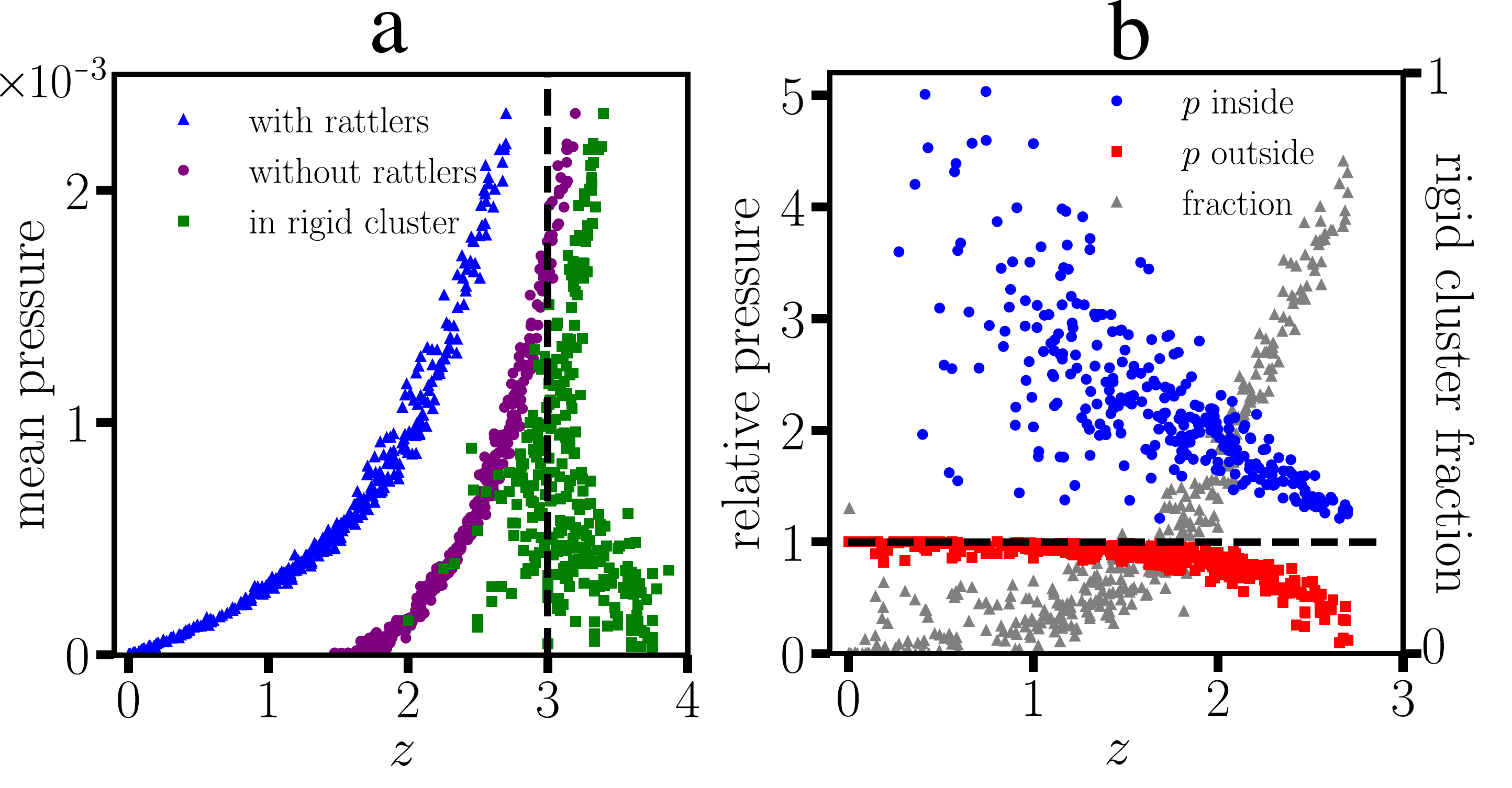}
	\caption{(a) Mean pressure $p$ as a function of $z$ of the entire packing (blue), within rigid clusters only (green), and with rattlers removed (purple). (b) Pressure inside rigid clusters (blue dots) and outside rigid clusters (red squares) normalized by the mean pressure of the entire packing. The fraction of rigid clusters (grey triangles) is also plotted for reference. }
	\label{fig:pressure}
\end{figure}

To show that the rigid clusters are mechanically relevant, we calculate the virial pressure $p$ from the contact forces (see \cite{SM}). 
In Fig.~\ref{fig:pressure}a, we show $p(z)$, rattlers included: the blue curve rises gradually but $z$ remains well below $3$. For comparison, we also include the same data but with the rattlers removed, as is commonly plotted \cite{majmudar2007jamming}; this curve now crosses $z=3$. 
In contrast, when $(p,z)$ are calculated using only bonds within the rigid clusters, we observe that $z \geq 3$  except in some very small clusters. 
In Fig.~\ref{fig:pressure}b, we compute the local pressure inside vs. outside the rigid clusters, normalized by $p$ for the entire packing. We find that pressure within rigid clusters is always significantly higher than the mean pressure. In contrast, the pressure in the floppy regions is always below average and drops further for $z \gtrsim 2$, while the mean pressure, the rigid cluster fraction, and the rigid region fraction all start to rise. We interpret Fig.~\ref{fig:pressure} as an emerging rigid backbone, responsible for the rise in pressure and carrying the majority of stress; this same mechanism was previously observed in simulations \cite{henkes2016}. 

\begin{figure}[t]
 \centering
 \includegraphics[width=\columnwidth]{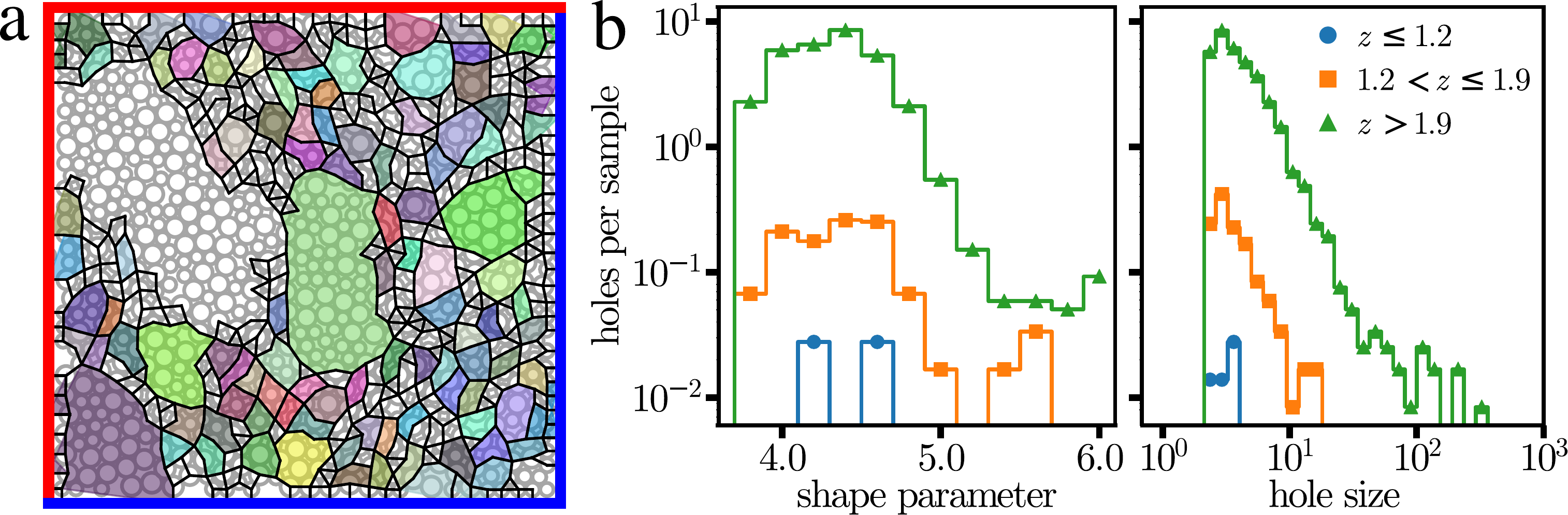}
 \caption{(a) Rigid clusters (black bonds) decomposed into tiles of closed loops; most colored tiles with $h>2$ contain non-rigid particles; $z=2.69$. (b) Histogram of hole sizes, in units of average particle area and histogram of hole shapes.}
 \label{fig:holes}
\end{figure}

In Fig.~\ref{fig:schematic}d, rigid clusters surround large \emph{holes} that contain floppy bonds and rattler particles. To characterize these floppy holes, we decompose the rigid cluster graph into a unique set of tiles, of which the larger ones correspond to the holes. Each tile corresponds to a face of the planar graph where the rigid bonds are the edges connecting vertices at the particle centers. To examine hole statistics, we employ a simple cutoff in hole size $h>2$, in units of mean particle area, to exclude (most) simple interstices between particles; the remaining tiles are colored in Fig.~\ref{fig:holes}a.
With increasing $z$, we observe both more and larger holes (Fig.~\ref{fig:holes}b), with  the system size as an apparent cutoff in hole size for $z>1.9$. We quantify changes in shape using the  dimensionless shape parameter $p_0 = P/\sqrt{A}$, where $P$ is the hole perimeter and $A$ is its area; a regular hexagon has  $p_0=3.72$ and larger values indicate less circular due to convexity changes and/or elongation. As $z$ increases, we observe a broader range of shapes with some jaggedness emerging. Thus, the rigid structures resemble a sponge-like porous medium much like the interior of sourdough bread. This finding is compatible with the presence of arch structures, rigid bridges, and hinges linking up rigid clusters to form a spanning network~\cite{liu2019}, and contrasts with the rigidity transition in frictionless packings, where such floppy holes are not observed \cite{ellenbroek2015rigidity}. 

{\it Discussion:} We have investigated the network structure of real, frictional granular materials under shear using two distinct, but compatible, measures of rigidity. 
We find a frictional jamming transition at $z_c= 2.4 \pm 0.1$, significantly below $z = 3$, the lower bound on stable frictional packings given by mean-field constraint counting and also known as random loose packing \cite{onoda1990}. Within the constraints of small system size and limited statistics, we observe a rigid cluster size distribution consistent with a continuous rigidity transition. 
Our $z_c$ is also lower than simulation results by \cite{Vinutha2019} who observed a rigid spanning cluster at $z_c \approx 2.9$ and the percolation of over-constrained bonds at $z=3$. Finally, our experimental results contrast with simulations modeling friction with rough, but frictionless particles, for which the transition occurs at the isostatic point~\cite{Papanikolaou2013,Wyart05}.

Open questions include what role the mechanics of these rigid structures plays in local failure under shear. While strong force chains often surround a floppy hole with an arch-like shape, we observed only partial correlation with pressure: not all forces within floppy regions are weak. Our results need to be complemented with observations of force chains~\cite{Forcechains1,Forcechains2,Nott} and cycles~\cite{Tordesillas} to more completely address rigidity in their descriptions. This could be done through topological~\cite{KarenTopology}, geometrical~\cite{Durian}, or stress-space approaches~\cite{Bulbul}. Identifying rigid structures will also be important for shear-thickening in dense granular suspensions, where a load bearing rigid cluster abruptly emerges via the exchance of frictionless, lubricated contacts for frictional contacts \cite{Wyart-Cates,Mari}. 
Our method provides a framework to go beyond mean-field in particulate systems, to ultimately understand the delicate interplay between constraints, forces, and geometry.
 

\paragraph*{Acknowledgments:} We acknowledge Arne te Nijenhuis for help with collecting the photoelastic data. We are grateful for support from
James S. McDonnell Foundation (K.E.D, J.E.K.), NSF grants
DMR-0644743 (apparatus), DMR-1507938 and DMR-1832002 (J.M.S.), and BBSRC grant  BB/N009150/1-2 (S.H.).

\paragraph*{Data and software availability:} The data and codes  will be available on DataDryad.org and GitHub \cite{PEGS,rigid}.

\let\oldaddcontentsline\addcontentsline
\renewcommand{\addcontentsline}[3]{}
\let\addcontentsline\oldaddcontentsline

\clearpage
\newpage

\begin{widetext}
\begin{center}
\textbf{\large Sponge-like rigid structures in frictional granular packings \\ Supplementary Material}
\end{center}
\end{widetext}

\setcounter{equation}{0}
\setcounter{figure}{0}
\setcounter{table}{0}
\setcounter{page}{1}
\makeatletter
\renewcommand{\thefigure}{S\arabic{figure}}
\renewcommand{\bibnumfmt}[1]{[S#1]}
\renewcommand{\citenumfont}[1]{S#1}

\appendix

\tableofcontents

\section{Robustness of contact identifications}

\begin{figure}[h]
	\begin{center}
		\includegraphics[width=\linewidth]{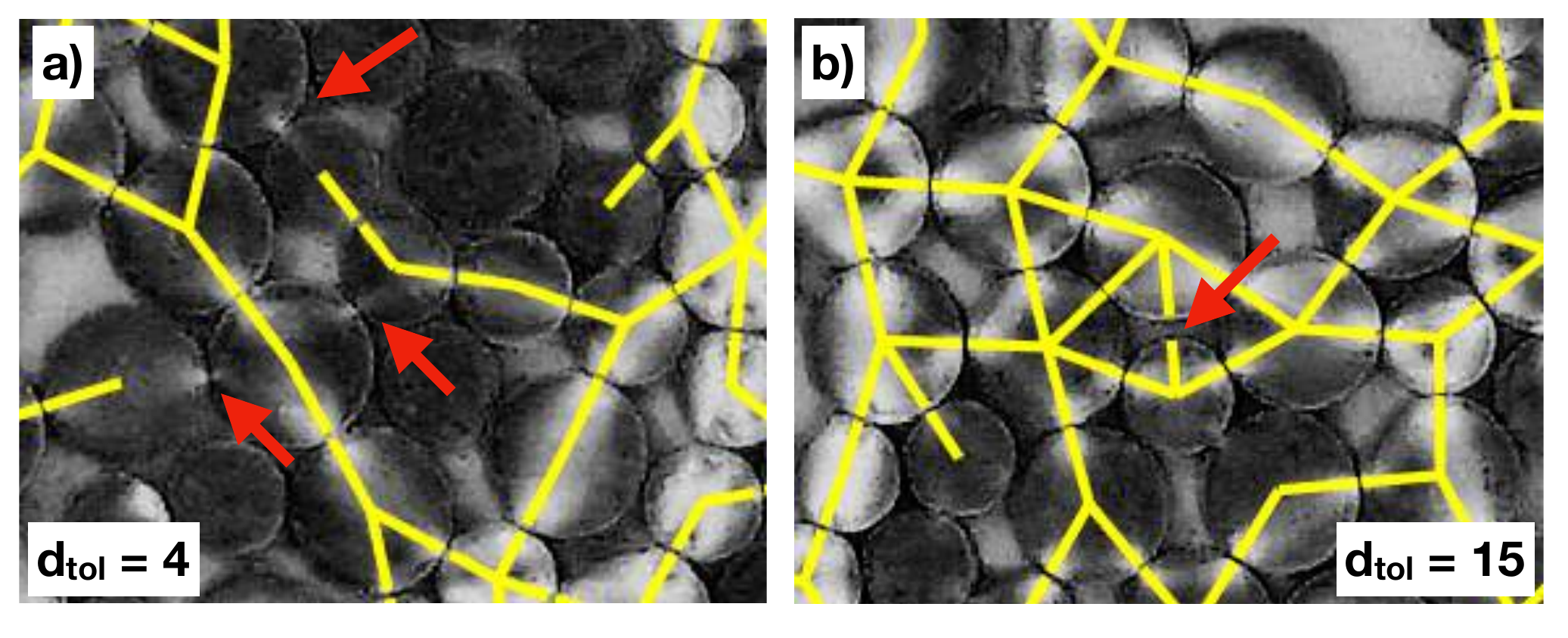}
	\end{center}
	\caption{Example images showing the under- and over-detection of contacts, as a result of a (a) too-strict or (b) too-lenient detection threshold $d_\mathrm{tol}$. Bright areas show a photoelastic response from the particles and yellow lines indicate identified contacts. False-negative and false-positive contacts are highlighted by red arrows.}
	\label{fig:visual}
\end{figure}

Any analysis of a contact network depends crucially on having an accurate determination of whether or not two particles are in contact. Here, we describe how we (1) validated our contact detection and (2) estimated uncertainties for measurements of the mean contact number $z$. Fig.~\ref{fig:visual} displays examples of under- and over-detection of contacts. 

\subsection{Sensitivity to parameter choices}

\begin{figure}
\begin{center}
    \includegraphics[width=\columnwidth]{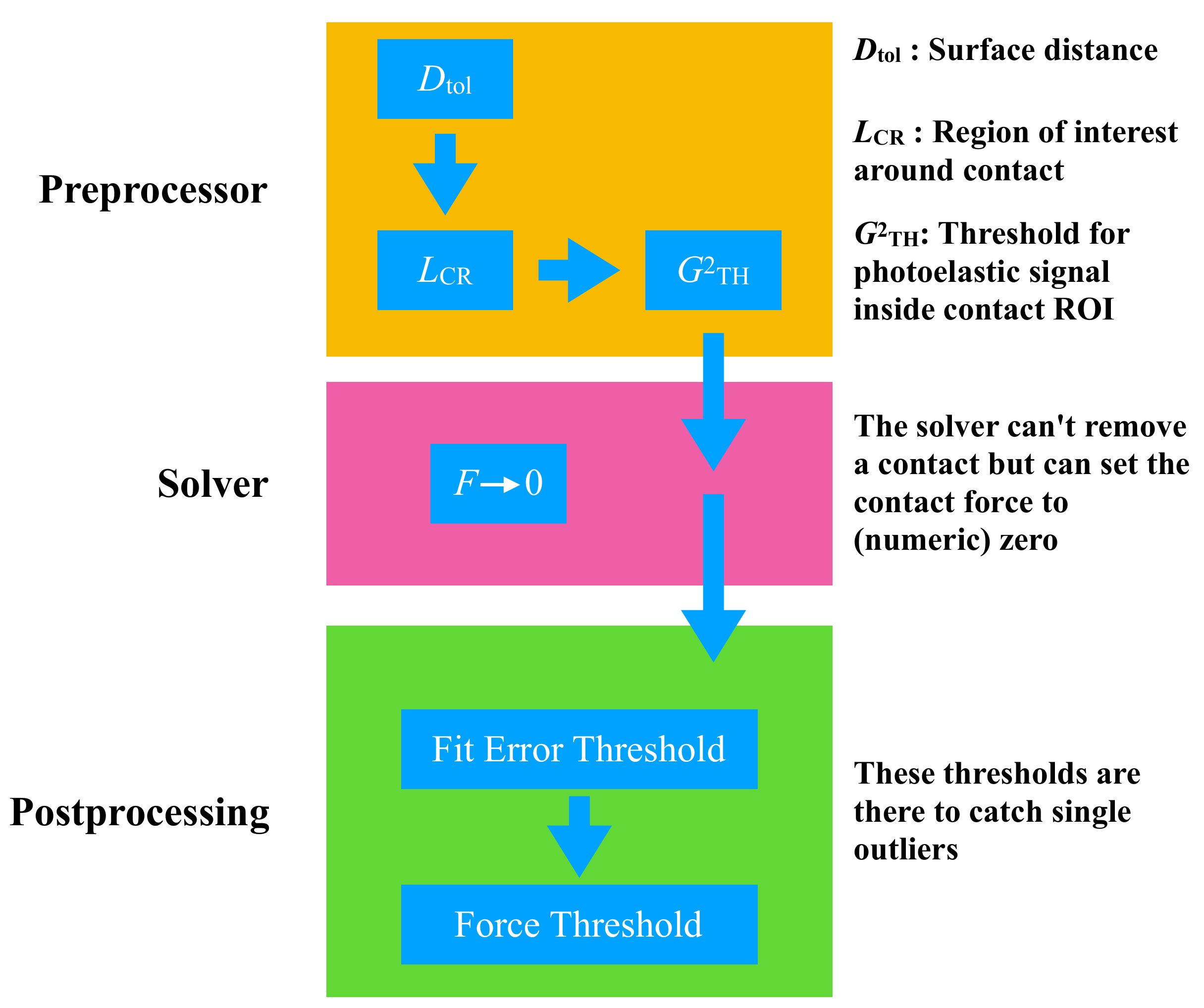}
\end{center}
    \caption{Contact processing pipeline. }
	\label{fig:pipeline}
\end{figure}

The open-source PeGS software package \cite{S-PEGS,S-daniels_photoelastic_2017} consists of three main parts, the preprocessor, the solver, and the postprocessor, as shown schematically in Fig.~\ref{fig:pipeline}. Each of these parts influences what is counted as a {\it valid contact} in the paper's analyses.  

\begin{figure}
\begin{center}
    \includegraphics[width=0.8\columnwidth]{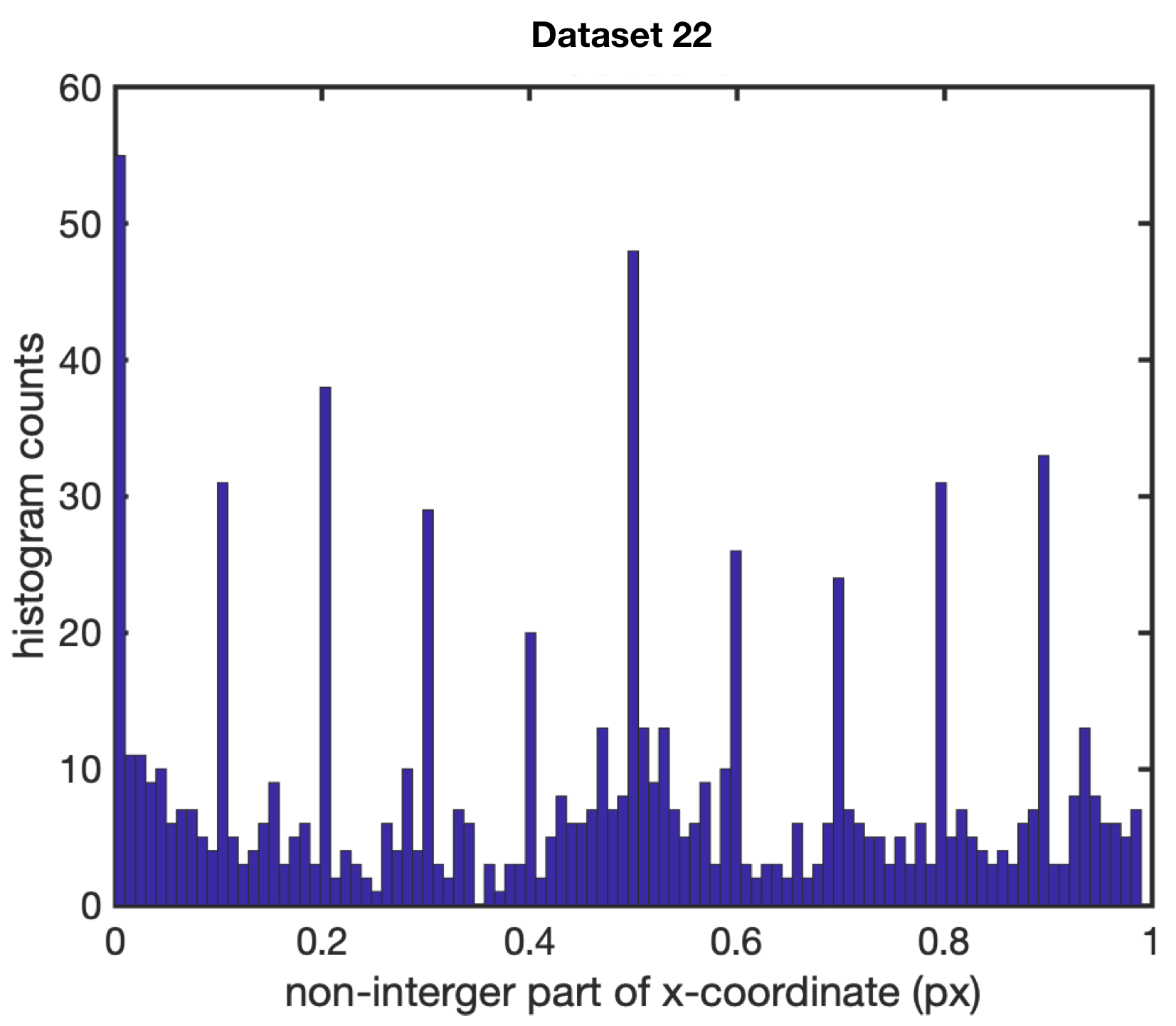}
\end{center}
    \caption{Histogram of noninteger parts of detected particle-coordinates show distinct peaks at intervals of 0.1 pixels. This value sets the precision of the center-detection of the particles.}
	\label{fig:subpixel}
\end{figure}

\medskip
The {\bf preprocessor} detects the particle locations and radii using a Hough transform. As shown in  Fig.~\ref{fig:subpixel}, our resolution is 0.1 pixels,  or approximately $R_1/200$, where $R_1$ is the radius of the smaller particles.

For each possible pair of particles, we compute the surface distance as the difference between the measured distance between their centers and the sum of the two particle radii. If this difference is below a threshold $d_\mathrm{tol}$, that pair of particles is considered to have a \emph{possible contact}. The quantity $d_\mathrm{tol}$ is nonzero in order to account for uncertainty in the particle positions and radius detection, not just particle deformation during contacts. For each possible contact, we measure the $G^2$ (gradient-squared) response \cite{S-daniels_photoelastic_2017, S-abed_zadeh_enlightening_2019} within a circular region of interest of radius $L_\mathrm{CR}$  pixels centered around the contact point. We set a threshold $G^{2}_\mathrm{th}$, above which the contact is accepted into the list of contact points for the packing.

Note that inclusion on this list is a necessary, but not sufficient, condition for being ultimately counted as a valid contact. It is still possible that the solver/postprocessor later determines that a contact has a undetectable level of force, and is discarded from the list. In the text that follows, we use the following values which we found to be appropriate for this particular experiment (combination of particle material and geometry, apparatus, lens, and camera): $d_\mathrm{tol}=0.5 R_1$, $G^2_\mathrm{th} = 0.15$,  and $L_\mathrm{CR}= 0.25 R_1$.  

\medskip
The {\bf solver} takes the list of possible contacts provided by the preprocessor, and uses a model \cite{S-PEGS,S-daniels_photoelastic_2017} of the  photoelastic response to determine, through optimization, the vector contact force at each contact. 
For every contact on the list, the solver also returns the residual $e$ between the fit result and the experimental data. A smaller value of $e$ indicates a higher-quality of the fit, to be evaluated during the postprocessing step (below).

\medskip
The {\bf postprocessor} creates the final list of valid contacts and forces in the form of an adjacency matrix. To be excluded from this list, two de-selection criteria apply:
\begin{enumerate}
	\item Force fit quality: if the residual $e > e_\mathrm{max}$, all forces from this particle are set to zero, effectively removing the contact from consideration. The value of $e_\mathrm{max}$ is set empirically to exclude obvious fit errors. 
	\item Force magnitude range: if the magnitude of the contact force falls outside a specified range  $F_\mathrm{min}< |F| < F_\mathrm{max}$, then that contact force is set to zero, effectively removing the contact from consideration. Here, we set $F_\mathrm{min} = 10^{-3}$~N and $F_\mathrm{max} = 2$~N, where the lower limit corresponds to an undetectable force and the upper limit is the force at which we are no longer able to properly resolve the photoelastic fringes.
\end{enumerate}

Setting these 3 values ($e_\mathrm{max},\,F_\mathrm{min},\,F_\mathrm{max}$), requires tuning by looking at the images and the output under the particular lighting conditions used in that experiment. Through visual inspection between what the algorithm detects and what is visible in the camera image, we choose an acceptable set of values for each of these three parameters.

\subsection{Robustness tests}

To determine the uncertainty in our contact detection, we conducted a robustness test to examine the sensitivity of contact detection to the choice of each of the parameters describe above. We tested these sensitivities for the preprocessor and postprocessor steps in isolation, and then performed a combined test to evaluate the ability of the solver to recover from poor-quality preprocessor data. We performed these tests on images from two different initial particle configurations (named Dataset 22 and Dataset 26), each under three different states of shear stress, for a total of 6 evaluations.

\begin{figure*}
\begin{center}
    \includegraphics[width=\textwidth]{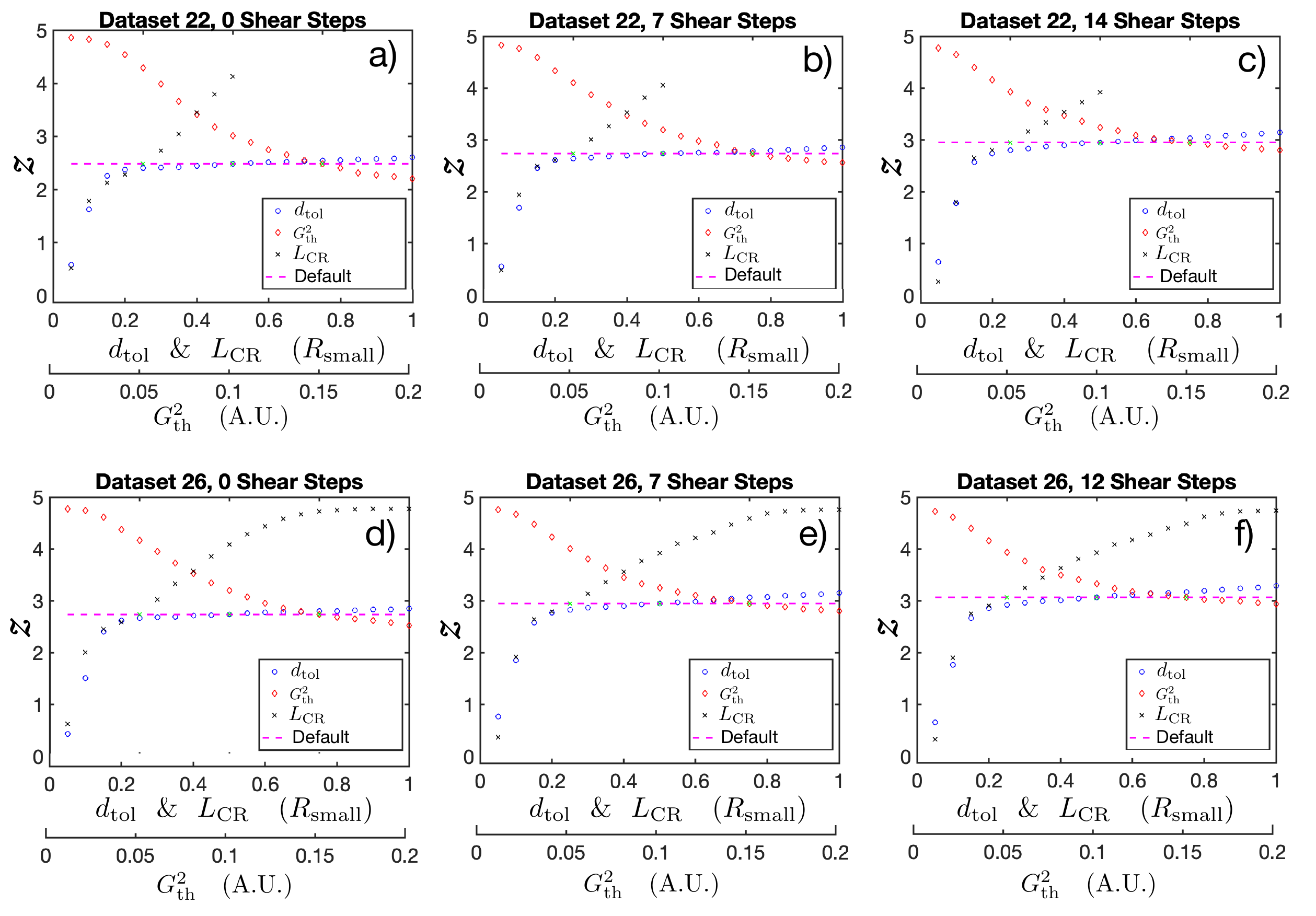}
\end{center}
    \caption{{\it Preprocessing robustness test:} The horizontal axis corresponds to varying $0.1   R_1 \le d_\mathrm{tol} \le 1.0 R_1$, $0 \le G^2_\mathrm{th} \le 0.2 $ and $0.1 R_1 \le L_\mathrm{CR} \le 1.0 R_1$, where $R_1$ is the size of the small particles in the image, each varied independently (other parameters held at their default value). The dashed line is the result when all parameters are set at their default value. }
	\label{fig:pre}
\end{figure*}

\paragraph*{Preprocessing:} For each of the 6 representative images, we varied one of the three parameters $d_\mathrm{tol}, G^2_\mathrm{th}, L_\mathrm{CR}$, while fixing the other two to their default values, and measured the average contact number $z$ after completing only the preprocessor step. The results are shown in Fig.~\ref{fig:pre}. 

We observe a strong dependence of the preprocessor-identified contacts on the user defined thresholds. For  $d_\mathrm{tol}$ and $G^2_\mathrm{th}$, there is a clear asymptotic value, while for $L_\mathrm{CR}$ there are images for which no sensible measurement is obtained for $L_\mathrm{CR} > 0.5\,R_1$ as the search area now strongly overlaps with other possible contact areas in the same particle. The solution is to allow for false-positives (larger value of $L_\mathrm{CR}$), which will be removed at later steps.

At the start of a new photoelastic experiment, suitable ranges of values were found by running the preprocessor several times, adjusting each of the thresholds until the preprocessor-detected contact network visually matches the contact network seen in the photoelastic images. 
We found that $L_\mathrm{CR}$ is the most difficult threshold to set correctly, and was done by comparing the detection result to the photoelastic image, as in Fig.~\ref{fig:visual}. {\it From these analyses, we select the values $d_\mathrm{tol} = 0.5 R_1$, $G^2_\mathrm{th} = 0.15$, and $L_\mathrm{CR} = 0.25 R_1$.}

\paragraph*{Postprocessing:} Using the possible contact lists generated during the preprocessor tests, we ran the photoelastic solver \cite{S-daniels_photoelastic_2017,S-PEGS}, and performed the postprocessing steps by varying one of the three parameters  ($F_\mathrm{min}, F_\mathrm{max} $ and $e_\mathrm{max})$ at a time, while fixing the other ones to their default value, and measured the resulting average contact number $z$.  

\begin{figure*}
	\begin{center}
		\includegraphics[width=\textwidth]{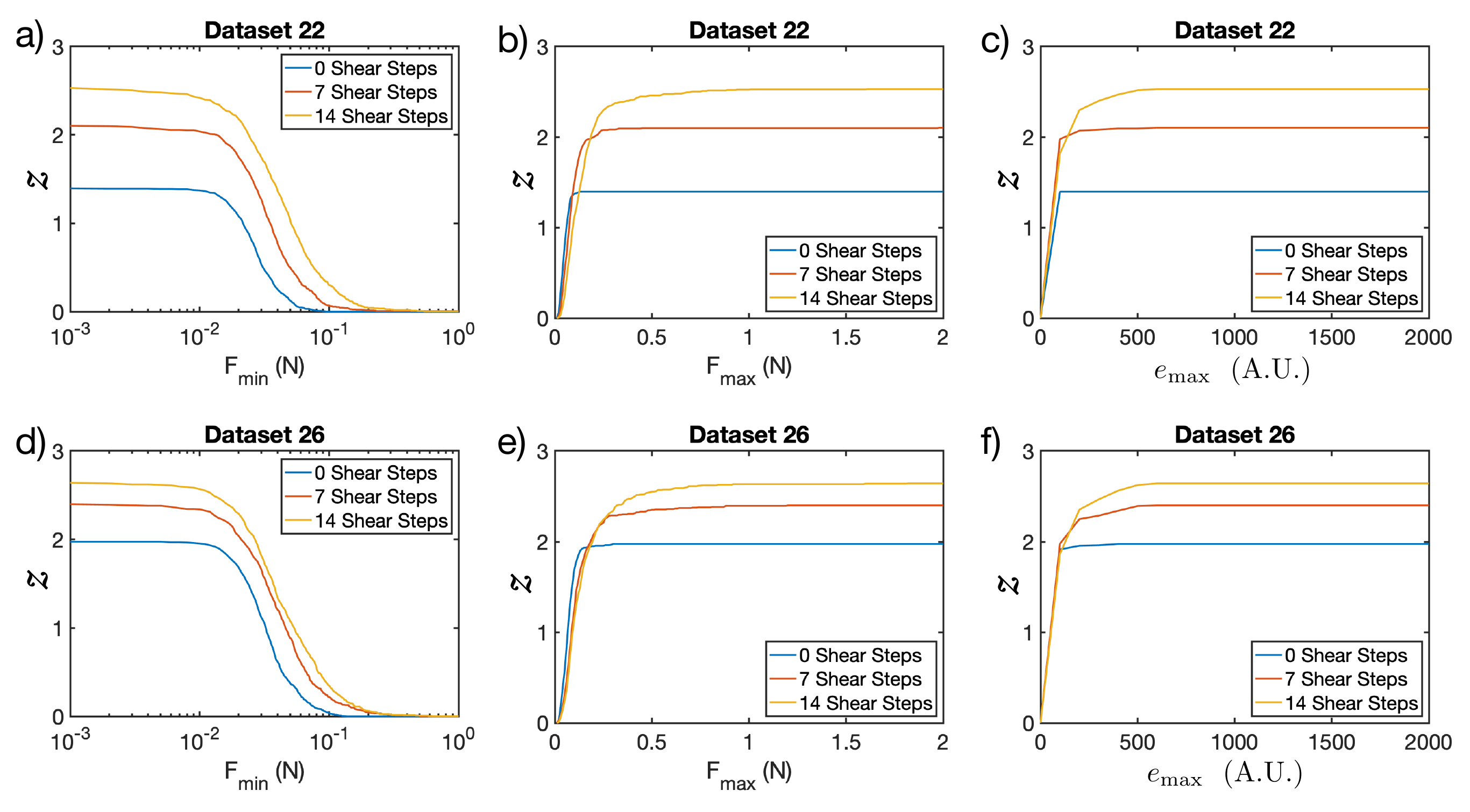}
	\end{center}
	\caption{{\it Postprocessing robustness test:} The horizontal axis corresponds to varying $10^{-3} \,\mathrm{N} \le F_\mathrm{min} \le 1.0 \,\mathrm{N}$, $10^{-2} \,\mathrm{N} \le F_\mathrm{max} \le 2.0 \,\mathrm{N} $ and $0 \le e_\mathrm{max} \le 2000$, each varied independently (other parameters held at their default value). The different coloured lines correspond to different loads on the system.}
	\label{fig:post}
\end{figure*}

As shown in Fig.~\ref{fig:post}, we observe that the mean contact number $z$ does not change significantly with the choice of threshold, except for unrealistically low values of $e_\mathrm{max}, F_\mathrm{max})$. The aim in selecting $F_\mathrm{min}$ is to exclude contacts that the solver set to a zero value. As can be seen in Figs. \ref{fig:post} (a) and (d) there is little change to $z$ for $F < 10^{-2}$~N so we set $F_\mathrm{min} = 10^{-3}$~N as our default value.

We observe that $z$ decreases from $2.5$ to $3.0$ after preprocessing to $1.5$ to $2.5$ after postprocessing, depending on which of the 6 images is considered. 
The threshold rules, given above, required that only a reduction in the number of contacts is possible. 
{\it From these analyses, we set the values $F_\mathrm{min} = 10^{-3}\, N$, $F_\mathrm{max} = 2 \,N$, $e_\mathrm{max}=2000$.}

\paragraph*{Solver:} 
To determine the uncertainties in $z$, we examined the variability as a function of the entire data pipeline, including the solver, for variable values of $d_\mathrm{tol}$. This tests the robustness of the process to the inclusion of false positives in the original contact list.

\begin{figure*}
	\centerline{\includegraphics[width=0.66\linewidth]{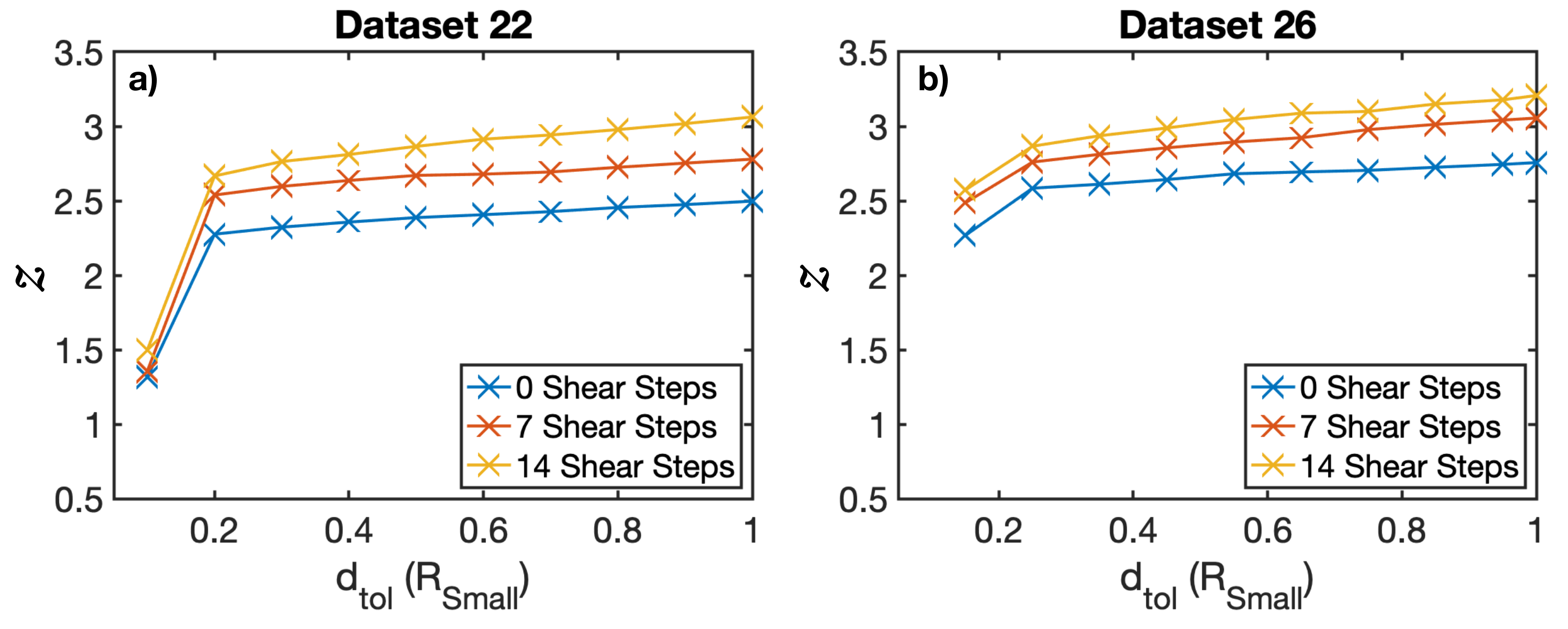}}
	\caption{{\it Solver robustness test:} Measured $z$ as a function of the choice of  $d_\mathrm{tol}$ for the preprocessor; all other parameters are the default values. We estimate the uncertainty in $z$ from the slope of  $z(d_\mathrm{tol})$. The different coloured lines correspond to different loads on the system.}
		\label{fig:solve}
	\end{figure*}

As shown in Fig.~\ref{fig:solve}, for all 6 images there is only a slow dependence of $z$ as a function of $d_\mathrm{tol}$ once a value of $d_\mathrm{tol} < 0.25 R_1$ is surpassed. Above that value, there is a systematic error in $z$ which is equally present in all datasets: the presence of false positives.

This result is consistent with visual observations, demonstrated by the two examples in Fig.~\ref{fig:visual}. In comparing the full set, we observed an optimum at  $d_\mathrm{tol} \sim 0.5 R_1$. For $d_\mathrm{tol} < 0.25 R_1$, the preprocessor missed obvious contacts (which cannot be recovered), and for $d_\mathrm{tol} > 0.75 R_1$ it selected physically impossible contacts (some of which, but not all, were later trimmed by the postprocessor). Thus, it is better to err on the side of the largest $d_\mathrm{tol}$ that works for all datasets (see Fig.~\ref{fig:pre}), and this is consistent with a choice of $d_\mathrm{tol} = 0.5 R_1$ . Our estimate of the uncertainty $z$ is, therefore, half the change across a reasonable set of choices for $d_\mathrm{tol}$, averaged over all 6 example images, which leads to an uncertainty in $z$  of $\pm 0.1$.

\clearpage
\section{Rigid clusters: the pebble game}

\subsection{Pebble game algorithm}

A rigid cluster is defined as the set of connected rigid bonds in a network that are mutually rigid with respect to each other.  A rigid cluster with no redundant bonds is a minimally rigid cluster.  Generically, a connected cluster (or network) with $N$ vertices is minimally rigid in two dimensions if and only if it has $2N - 3$ bonds and no subcluster of $n$ vertices has more than $2n-3$ bonds, which is Laman's theorem~\cite{S-laman70}.  This theorem is applicable to two-dimensional systems with central force constraints, such as bar-joint networks or frictionless packings described in terms of the contact network, with $2N$ denoting the number of degrees of freedom and $3$ denoting the number of trivial, global zero modes, i.e. two translations and one rotation. The straightforward three-dimensional central-force extension of Laman's theorem does not rigorously hold~\cite{S-Servatius}. However, mathematicians have been able to characterize the generic rigidity of rigid bodies connected by bars, i.e. body-bar rigidity, in arbitrary dimensions via a tight $(k,k)$ network containing $N$ vertices and $M$ bonds such that every subset of $N'\le N$ vertices connects with at most $kN'-l$ bonds and $M=kN-l$~\cite{S-Tay1984}.

\begin{figure}[t]

\centering

\includegraphics[width=0.96\linewidth]{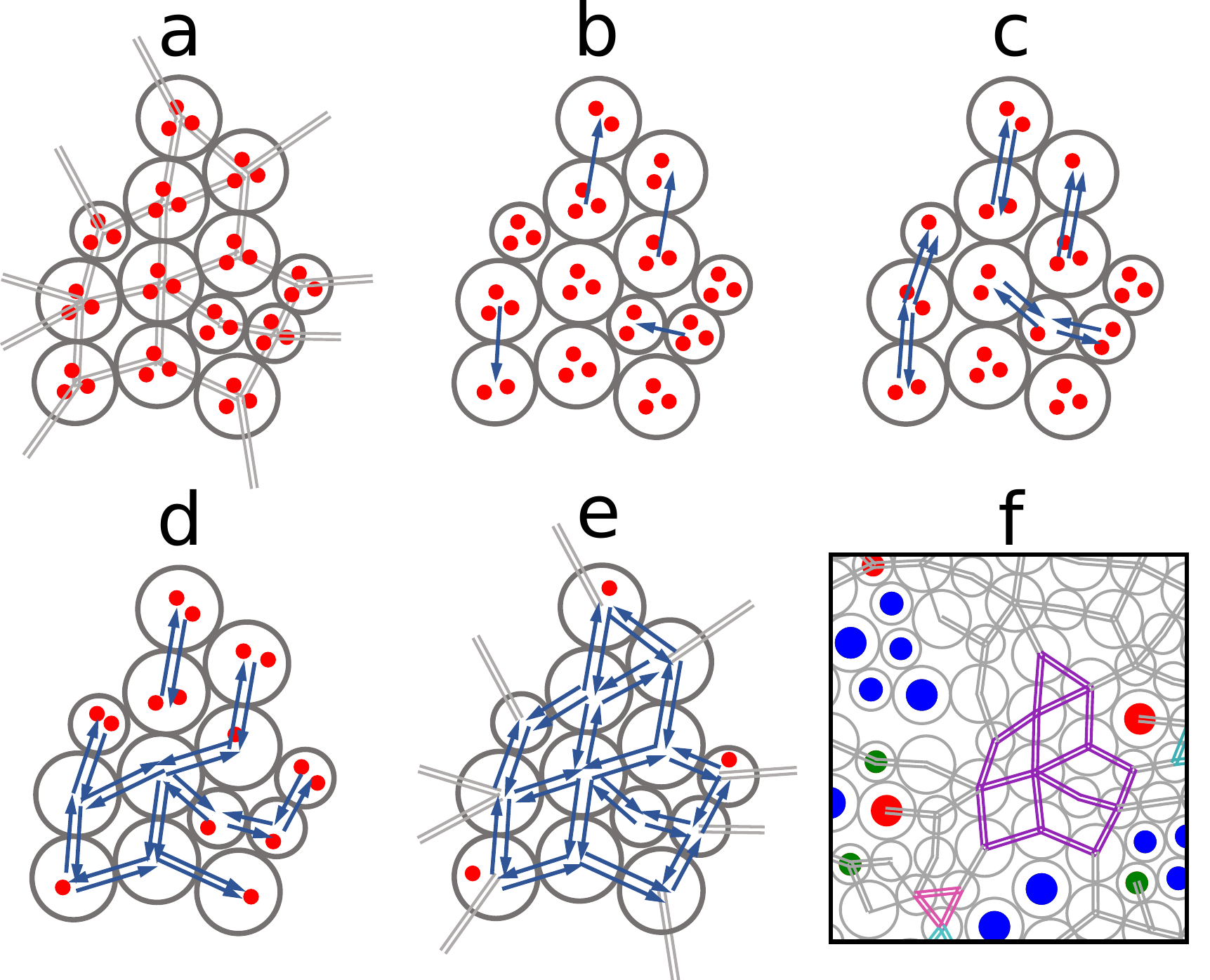}  \caption{Implementation of the pebble game for a subset of a packing (a)-(e) and the corresponding rigid cluster (f). }
\label{fig:pebble_game_flow}
\end{figure}

The \emph{ (k,l) pebble game}~\cite{S-jacobs1997algorithm,S-Lee2008} provides a combinatorial algorithm for determining which bonds in a network are rigid, from which the rigid clusters can then be determined. The integer $k$ represents the number of degrees of freedom for each particle, and the positive integer $l$ represents the number of global degrees of freedom for the system. 
The original $(2,3)$ pebble game by Jacobs and Hendrickson is applicable to frictionless packings~\cite{S-jacobs1997algorithm} which have two translational degrees of freedom per particle, while the extension to general $(k,l)$ was developed by Lee and Streinu~\cite{S-Lee2008} and is the one relevant to frictional packings~\cite{S-henkes2016}.  We provide here a brief description of the algorithm:

Initially, $k$ pebbles are ``placed'' on each vertex of the network (Fig. \ref{fig:pebble_game_flow}a). These vertices represent the {\it constraint network} defined by the contacts between the particles.  To conduct the pebble game, there is an additional {\it directed network} constructed from the constraint network upon which the pebbles are moved around (Fig. \ref{fig:pebble_game_flow}b-e).  As pebbles are moved around, two rules must be obeyed: 

\begin{description}

\item[Rule 1] No more than $k$ pebbles can be present on any vertex.

\item[Rule 2] A directed bond is accepted into the directed network when at least a total of $l+1$ pebbles are present at the two vertices defining the bond.

\end{description}

As each bond of the (undirected) constraint network is considered in turn, testing to see whether its associated bond is accepted into the directed network, the valid moves along are: 

\begin{description}

\item[Move A] A pebble found via a depth-first search starting a vertex $x$ may be moved along the path with the arrows of the directed path reversed until reaching vertex $x$.

\item[Move B] If there is a directed bond is accepted into the directed network via Rule 2, then the found pebble is removed from the directed network. 

\end{description}

The pebble game is played until all bonds in the constraint network have been considered. If, for example, there are more than $l$ pebbles that have not been removed from the directed network, then the constraint network is floppy.

This algorithm ensures that the bonds  accepted into the directed network map to independent constraints in the constraint network, so if there are $l$ pebbles left over and bonds that have not been accepted into the directed network, then these correspond to redundant bonds.  Note that such bonds are not necessarily unique since they may change depending on the order in which the constraint network bonds are considered.

Now let us complete the application of the $(k,l)$ pebble game to frictional packings. In such systems, the translational and rotation degrees of freedom must both be considered. For a 2D frictional granular system, we play a $k=3$ pebble game, named for the two translational plus one rotational degree of freedom for each particle. The system itself has $l=3$ global degrees of freedom (two translational plus one rotational).

In order to assign bonds for the correct number of independent constraints in the constraint network, it is necessary to ascertain whether each contact is below, or at, the Coulomb criterion. For frictional contacts (contacts below the Coulomb threshold), the normal and tangential components are independent and a double bond is assigned to the constraint network. For sliding contacts, the tangential and normal forces are no longer independent, and a single bond is assigned to the constraint network.

Once the constraint network is constructed, we then play the $(3,3)$ pebble game for that constraint network. This has been done for both frictional packings \cite{S-henkes2016,S-liu2019} and rigid-beam \cite{S-Tay1984} networks, with the double-bond structure having different meanings between the two cases. This is illustrated in Fig.~\ref{fig:pebble_game_flow}. Note that there are specific gearing motions (even cycles) that are not correctly captured by the frictional $(3,3)$ pebble game~\cite{S-Lester2018}. As free gearing motions are only possible for systems without 3-cycles (particles in a triangle), they are vanishingly rare in disordered granular packings.  We have therefore argued previously that near the frictional jamming transition, the algorithm is reasonably proficient~\cite{S-liu2019}. This paper, in which we observe a very strong correlation with the rigid regions gives quantitative evidence of its suitability.

To account for boundaries in the experimental system, each of the 4 walls is treated as a boundary particle which has a bond with some of its neighboring particles. Since the boundary is not photoelastic, we determine which particles are in contact with it by examining all particles located less than $1.2 R_1$ from the boundary. For each candidate particle, we accept it as a particle in contact with a boundary particle if the vector sum of its forces is non-zero, within a tolerance. We include an additional normal force between the candidate particle and the boundary particle so that the vector sum of the candidate particle is zero and then determine whether the contact is included as a single or double bond using the Coulomb criterion.

Once the constraint network is formed, we perform the  $(3,3)$ pebble game on that set of bonds. Fig.~\ref{fig:pebble_game_flow} shows pictorially how the pebble game is implemented on a subset of an experimental packing with $\mu=0.3$.  

The pebble game for frictional packings is implemented in the \emph{Pebbles} class of the rigid analysis python library \cite{S-rigid}.

\subsection{Floppy hole analysis}

\begin{figure}[h]
 \centering
 \includegraphics[width=0.48\linewidth]{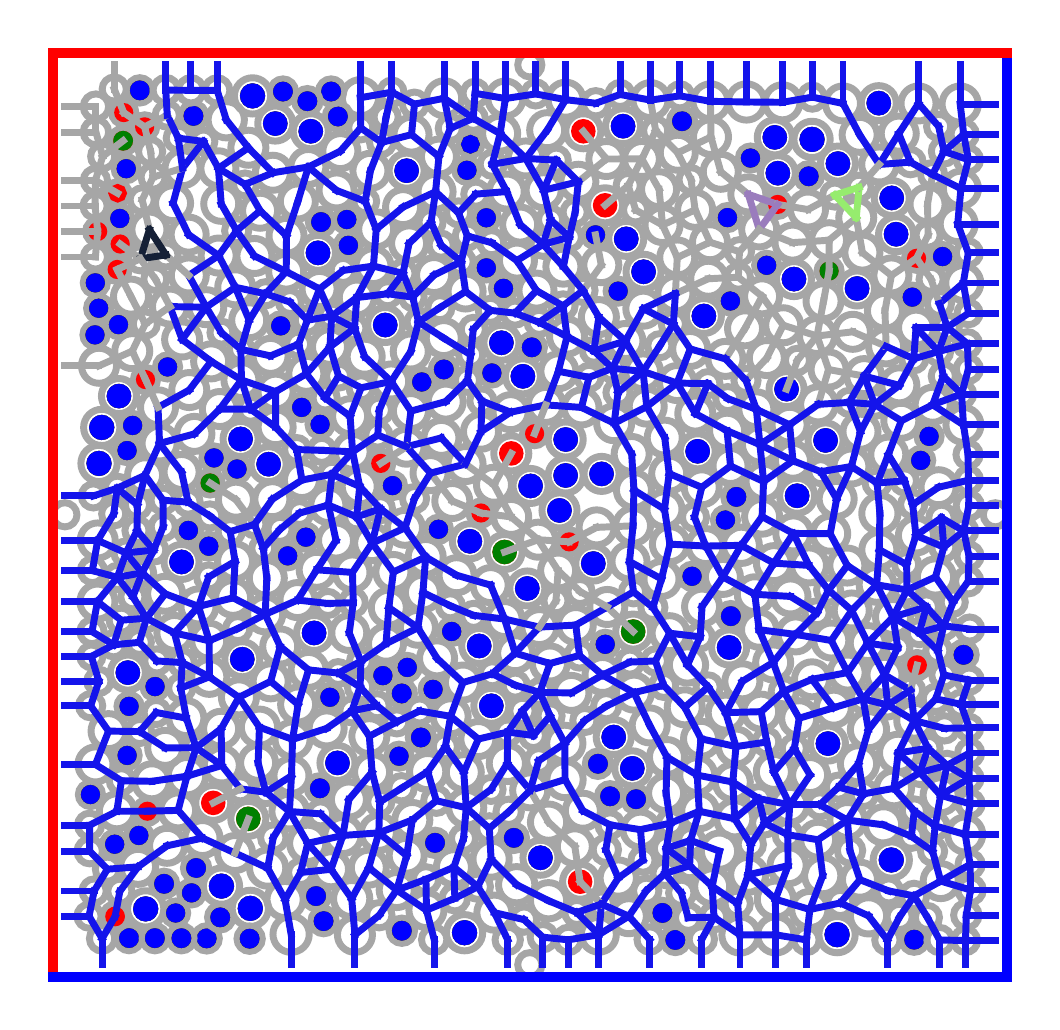}
 \includegraphics[width=0.48\linewidth]{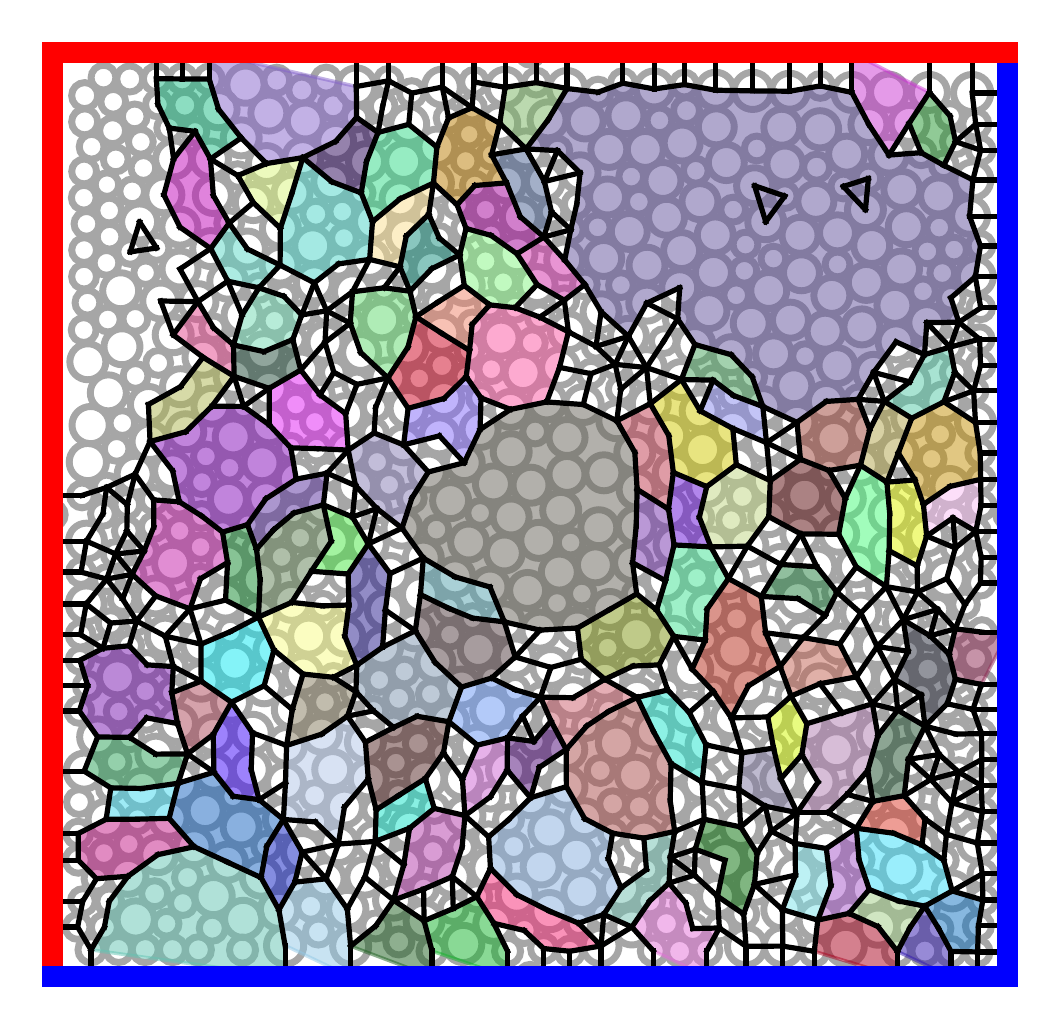}
 \caption{Rigid cluster plot and corresponding colored floppy hole decomposition of one sample with $z=2.58$. The irregularly shaped floppy regions outside of the rigid cluster are topologically not holes within the rigid cluster and do not form part of the analysis.}
  \label{fig:hole_sample}
\end{figure}

To analyse the structure of the rigid cluster, we identify \emph{floppy holes} containing rattler particles and floppy bonds in their interior. For example, the central hole in the rigid cluster of Fig.~\ref{fig:hole_sample} contains 22 particles, of which 9 are outright rattlers (marked blue), while the other ones are connected to each other and to the rigid cluster by floppy bonds (marked grey). To automatically identify such holes, we need to identify the faces of the planar graph associated to the rigid cluster (in blue in Fig. \ref{fig:hole_sample}a): the floppy holes simply correspond to the faces with a sufficiently large area to contain particles in their interior.

We construct a half-edge data structure by starting from a given bond and an orientation \emph{from} particle $i$ \emph{to} particle $j$, and then moving counterclockwise by selecting the \emph{next} bond from the bonds emanating from $j$ that makes the largest angle with $ij$ within the interval $(-\pi, \pi)$. Repeating this procedure moves counterclockwise around a face and will eventually close the loop by selecting $ij$ as the \emph{next} bond. Repeating this algorithm for all bonds and orientations that are not part of a face already allows us to identify the individual faces of the graph. 
When determining the faces, the effectively extended boundary particles are moved outward from the packing to avoid singular behaviour of the algorithm caused by intersecting contacts (i.e. a non-planar graph). 

To identify the faces that correspond to floppy holes, we first compute the mean area per particle by dividing area of the system by the total number of particles, or $A_0=\frac{L_x  L_y }{N}$, where $L_x$ and $L_y$ are the lateral dimensions of the region.  This will be the fundamental hole area unit. Then we compute hole area $A$ for all faces using the formula for a non-intersecting planar polygon. Empirically, a hole with size $A/A_0$ smaller than $2$ is not big enough to contain at least one particle in it, and we define floppy holes to be faces with $A/A_0\ge 2$. The faces identified as floppy holes for the sample in Fig.~\ref{fig:hole_sample}a are shown in color in Fig.~\ref{fig:hole_sample}b, including the central hole in a muddy color.

When compiling the hole statistics as shown in Fig.4b-c in the main text, we normalise $n_h$ such that the integral under each curve is equal to the mean number of holes per sample. We also compute the shape of the floppy holes using the dimensionless shape parameter $p_0=P/\sqrt{A}$, where $P$ is the floppy hole perimeter and $A$ is its area.  In Fig. 4c of the main text, we plot floppy holes per sample as a function of shape parameter for different $z$. 

\clearpage
\section{Rigid regions: the dynamical matrix}

\subsection{Frictional equations of motion}
We invoke the well-established Cundall-Strack model~\cite{S-Cundall-Strack} as a reasonable model for the equations of motion for our frictional experimental system. This is the same model that was used to establish the frictional pebble game \cite{S-henkes2016}. The equations of motion for individual particles are
\begin{align*}
m_i \ddot{\mathbf{r}}_i &= \mathbf{F}_i + \sum_{j\:n.\:i} \mathbf{F}_{ij} \\
I_i \ddot{\theta}_i &= T_i + \frac{1}{2} \sum_{j\:n.\:i} \mathbf{r}_{ij} \times \mathbf{F}_{ij},
\end{align*}
where $m_i$ is the mass of particle $i$ and $I_i$ is its moment of inertia, and the sums are over particles in direct contact. Here the first equation is for the forces, and the second is for the torques; we have made the approximation that force moments apply at the middle of the lever arm connecting particles.

\medskip
The pair forces can be decomposed into three contributions:
\begin{enumerate}
\item The elastic central forces due to deformation of the cylindrical particles, $\mathbf{F}_{ij}^{el} = - K_n (R_i+R_j -|\mathbf{r}_{ij}|) \hat{\mathbf{n}}_{ij}$, where $K_n$ is the (vertically integrated) Young's modulus of the particles. Note that such a \emph{harmonic} force is appropriate for cylinders with a continuous contact line as in the experiment, while \emph{Hertzian} forces are appropriate for a contact point.
\item The frictional forces between particles, which in a tangential frictional loading scenario can be written in a differential form $d\mathbf{F}^{\mu}_{ij} = K_t dt$, where $dt$ is the tangential displacement (see below).
\item The viscous dissipative forces, which we assume are dominated by dissipation within the elastic material, and we estimate $\mathbf{F}^{v}_{ij} = -\zeta (\mathbf{v}_i - \mathbf{v}_j)$ for particles in contact, and $0$ otherwise.
\end{enumerate}
Finally, there are also some single-particle dissipative forces due to interaction with the air flow around the particles, which we roughly estimate as $\mathbf{F}^v_i = - \zeta_{\text{air}} \mathbf{v}_i$ and $T_i =  - \zeta_{\text{air}} R_i \dot{\theta}_i$.

\subsection{Effective potential for friction}

\begin{figure}[h]
 \centering
 \includegraphics[width=0.7\columnwidth]{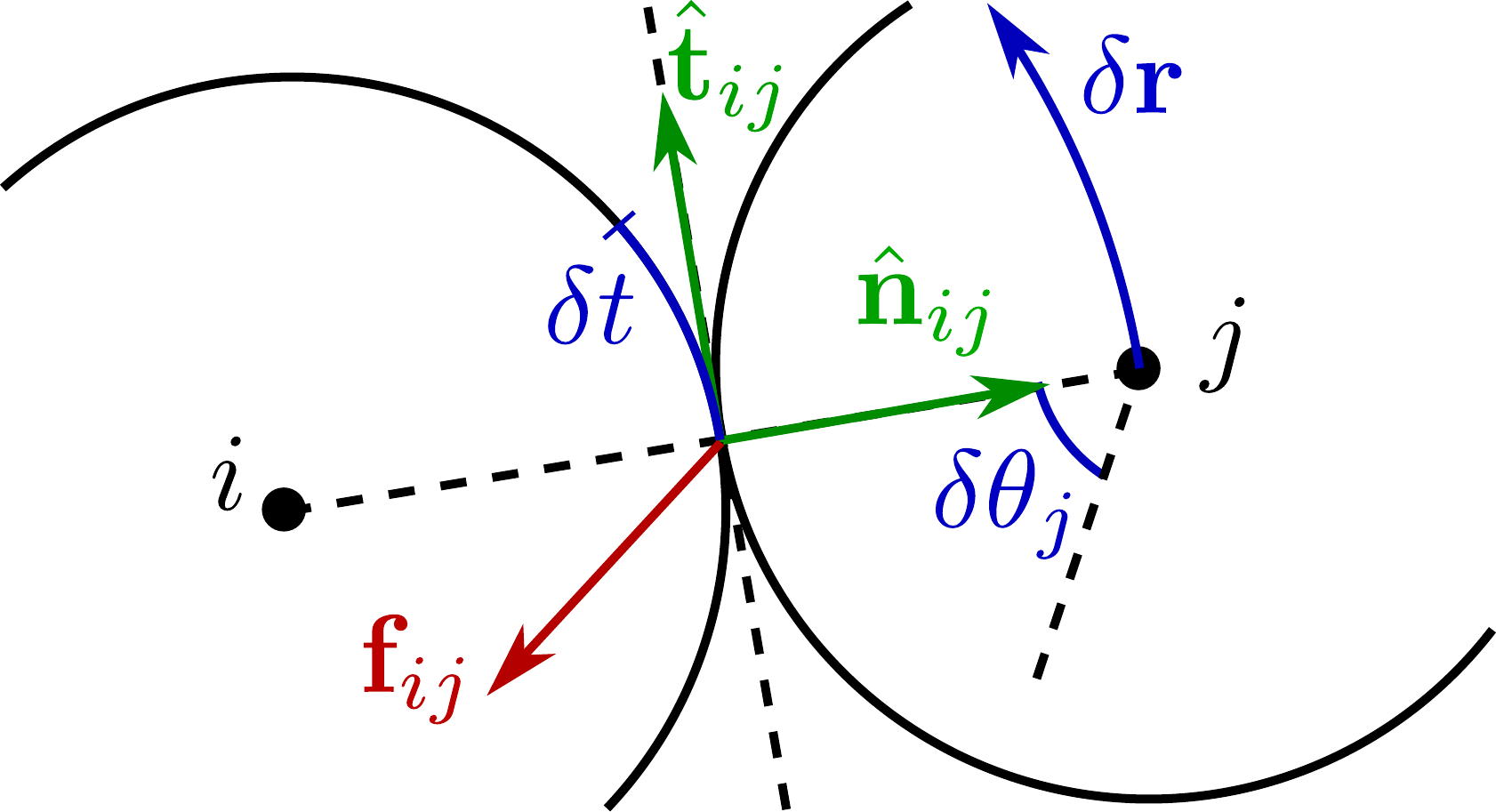}
 \caption{Local geometry around a frictional contact.}
 \label{fig:local_geometry_SI}
\end{figure}

In the limit of quasistatic deformations, the dissipative velocity-dependent terms will vanish, and the elastic and frictional forces will dominate. We therefore consider only the latter in our dynamical matrix approach. 
In the limit of small displacements (and only in this limit), we can recast the frictional forces as an effective potential. Consider the local geometry of a frictional contact shown in Figure \ref{fig:local_geometry_SI}. 

\begin{itemize}
 \item Particle $i$ at position $\mathbf{r}_i$ is joined to particle $j$ at position $\mathbf{r}_j$ through the contact $\mathbf{r}_{ij} = \mathbf{r}_j - \mathbf{r}_i$. 
 \item The normalised contact vector is the contact normal $\hat{\mathbf{n}} = \mathbf{r}_{ij}/|\mathbf{r}_{ij}|$.
 \item The tangential unit vector at the contact is $\hat{\mathbf{t}} = (n_y,-n_x)$ 
 \item The total force at the contact from $j$ on $i$ is $\mathbf{f} = -f_n \hat{\mathbf{n}} + f_t \hat{\mathbf{t}}$, where we have chosen $f_n>0$, i.e. directed towards $i$.
\end{itemize}

Translations are parameterized by $\delta \mathbf{r} = \delta \mathbf{r}_j - \delta \mathbf{r}_i$, while rotations are initially parametrized  by the two angular displacements $\delta \theta_i$ and $\delta \theta_j$. However, inspecting Fig.~\ref{fig:local_geometry_SI} reveals that the amount of tangential sliding \emph{at} the contact is determined by both rotations and translations, which can be written as 
\begin{equation} \delta t = \delta \mathbf{r} \cdot \hat{\mathbf{t}} - (R_i \delta \theta_i + R_j \delta \theta_j). \end{equation}  
To better understand this expression, here are two illustrative examples.  First consider two equal-sized particles in a gearing motion. In this case, $\delta \theta_i = - \delta \theta_j$ and $\delta t = 0$. Second, consider now a purely tangential translation of particle $j$ while $i$ remains fixed, for which the tangential sliding is $\delta t = |\delta \mathbf{r}_j|$. In this second case, there must be a compensating gearing motion of the second particle, so that we again have $\delta t = 0 $ if $R_j \delta \theta_j = |\delta \mathbf{r}_j|$.

\begin{figure}[h]
 \centering
 \includegraphics[width=0.7\columnwidth]{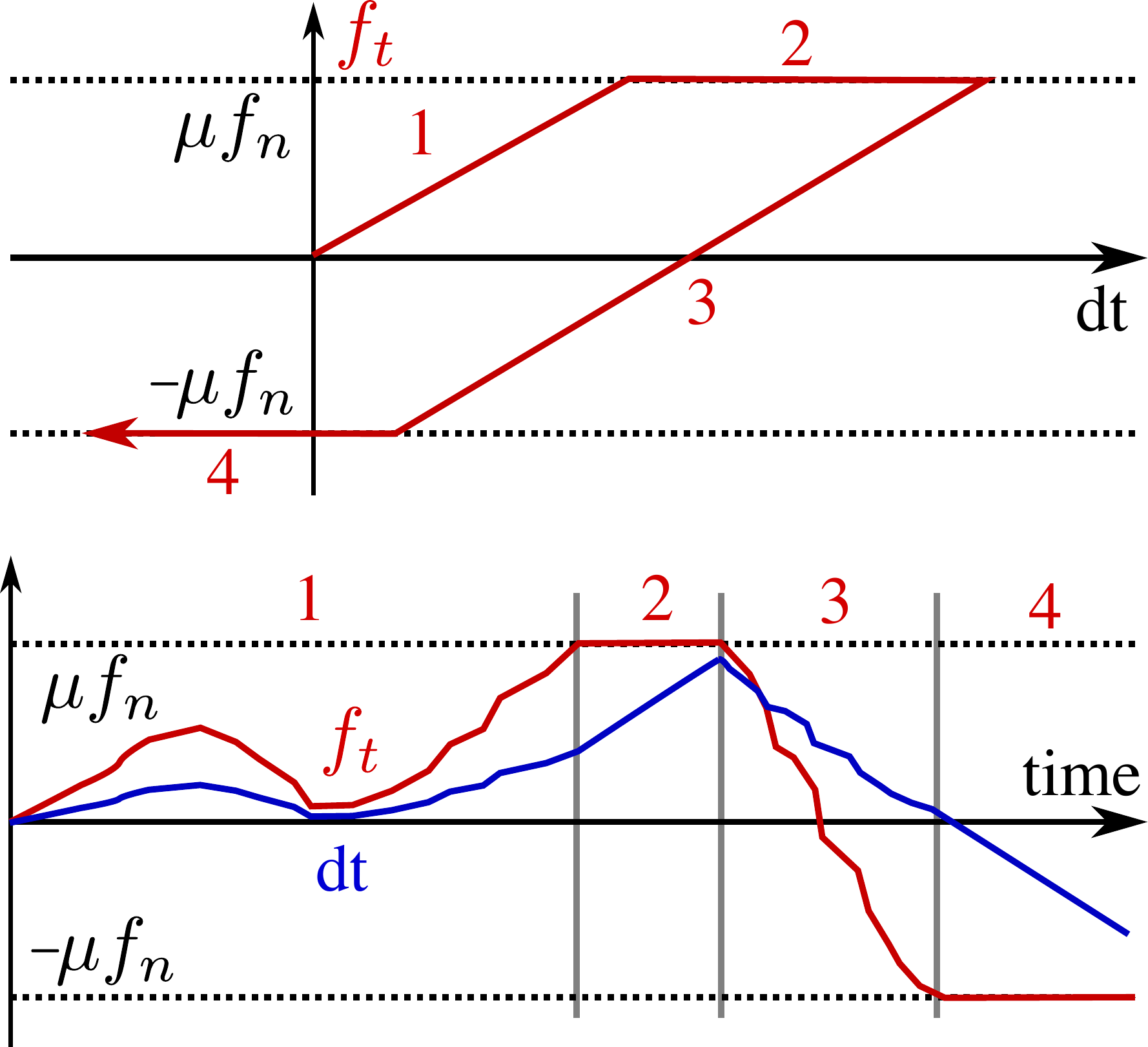}
 \caption{Schematic evolution of a frictional contact. Top: Magnitude of the frictional force as a function of tangential sliding coordinate. Bottom: Magnitude of the frictional force and amount of tangential sliding as a function of time.}
 \label{fig:friction_force_SI}
\end{figure}

We can now carefully consider the effect of a frictional force. In Figure \ref{fig:friction_force_SI}, we schematically show the evolution of a representative frictional contact. At time $0$, the contact is made by particles $i$ and $j$ coming sufficiently close. For simplicity, we will assume that $f_n$, the magnitude of the normal force, is approximately constant at short times; a reasonable assumption for a dense, slowly sheared packing. Upon contact, in phase 1 (tangential force loading), the increment in frictional force is given by $\delta f_t = K_t \delta dt$, where $\delta dt$ is the the infinitesimal amount of frictional tangential loading. We can also directly write $f_t = K_t dt$, during this phase tangential motion is reversible. In phase 2, the contact hits the Coulomb criterion $|f_t| =\mu f_n$, and then continues sliding, where for simplicity, we have assumed that the static and dynamic friction coefficients are the same. During phase $2$, the tangential sliding coordinate $dt$ continues to increase. In phase 3, the contact motion reverses, so that we again have $\delta f_t = K_t \delta dt$, and the frictional force falls below the Coulomb threshold. If the frictional force hits $f_t = - \mu f_n$, we enter phase $4$ where the contact slides in the opposite direction and $dt$ continues to decrease.

We can write an effective potential in linear response for phases 1 and 3, if we neglect the likelihood of transitioning between phases during an infinitesimal displacement. The only complicated situation that potentially arises regularly is when a sliding contact reverses direction, i.e. during the transition from phase 2 to phase 3. As tested in our simulations of \cite{S-henkes2016} (unpublished), for slowly sheared systems where the motion is primarily in a single direction, this occurs extremely rarely. This will, however, generate a major source of hysteresis when the shear is reversed. In our experiments, we do not take data at that instant of reversal, but only after the shear step has already occurred, putting us into phase 3 or 4.

During phase $1$, we can write an effective potential
\begin{equation} V_\text{eff}^f = \frac{1}{2} K_t dt^2. \end{equation}
We can see this by taking the explicit gradient of the potential in the local $\hat{\mathbf{n}}$, $\hat{\mathbf{t}}$ coordinate system, with coordinates $\delta n = \delta \mathbf{r} \cdot \hat{\mathbf{n}} $ and $\delta t =  \delta \mathbf{r} \cdot \hat{\mathbf{t}} $. Note that $dt$ is a scalar quantity, so despite $\hat{\mathbf{n}}$ and $\hat{\mathbf{t}}$ being a moving frame, the covariant derivative is the same as the ordinary derivative. Therefore, the effective potential remains valid for finite displacements and we have
\begin{align*}
\mathbf{F}_t &= - \nabla_{\mathbf{r}_i} V_\text{eff}^f   = - K_t dt \nabla_{\mathbf{r}_i} dt \\
& = - K_t dt \left [ \frac{\partial dt}{\partial \delta n} \hat{\mathbf{n}} + \frac{\partial dt}{\partial \delta t} \hat{\mathbf{t}} \right] = -K_t dt  \hat{\mathbf{t}}. 
\end{align*}

During phase 3, the same potential applies with a shift, $V_\text{eff}^f = \frac{1}{2} K_t (dt-dt_r)^2$, where $dt_r$ is the tangential displacement at the moment of the transition between phases $2$ and $3$. During phases $2$ and $4$, the tangential force is a constant, and the effective potential is simply $V_\text{eff}^f =\pm \mu f_n dt$.

\subsection{The dynamical matrix with friction}

The dynamical matrix provides the equations of motion for a particle within its local potential, as has traditionally been done for analyzing the vibrational modes of crystals~\cite{S-AshcroftMermin}. In the context of jammed systems~\cite{S-van_hecke_2009,S-liu_jamming_2010}, rigid regions are constructed by locating all bonds for which the relative motion of the two connected particles falls below a threshold value, and is thereby treated as zero: that contact is said to be rigid. Even though frictional particles are governed by nonconservative forces, we can nonetheless calculate an effective dynamical matrix using our effective potential. Following the method that was only outlined in Refs.~ \cite{S-henkes2010, S-somfai2007critical}, we begin by expanding the equations of motion for each particle $i$ about its equilibrium position to obtain 
\begin{align}
\delta\ddot{r}^i_{\alpha\beta} &=-D^{ij}_{\alpha\beta}\delta r^j_{\beta}+\text{dissipation}(\delta \dot{r}) + O(\delta r^2), \\
 D^{ij}_{\alpha,\beta} & =  \frac{1}{\sqrt{m_{i,\alpha} m_{j,\beta}}}\frac{\partial^2 V_{ij}}{\partial r_{i,\alpha} r_{j,\beta}}. \nonumber
\end{align} 
Here $D^{ij}_{\alpha,\beta}$ is the dynamical matrix of the system, and the indices $(i,j)$ label all disks, while $(\alpha,\beta)$ labels the two spatial $x,y$ components and the angular component $\delta\theta$,  and $m$ denotes the particle mass or the moment of inertia depending on the type of component. 

Using this framework of relative motions, we can then write the linearized interparticle potential around the contact as
\begin{equation}
 V_{ij} = \frac{1}{2} \left[ K_n (\delta \mathbf{r} \cdot \hat{\mathbf{n}})^2 - \frac{f_n}{|\mathbf{r}_{ij}|} \left( \delta \mathbf{r} \cdot \hat{\mathbf{t}}\right)^2 + \delta V_{\text{eff}}^f \right],
\end{equation}
where the last term is the effective frictional potential rewritten for an infinitesimal displacement, $V_{\text{eff}}^f = K_t \delta t^2$ for phase $1$ and $3$, i.e. a loading contact (stick), and $V_{\text{eff}}^f  = \pm \mu f_n \delta t$ for a sliding contact.
The first term is simply the spring potential responsible for the elastic normal forces. The second (negative) term might seem counterintuitive: it comes from the existing normal force at the contact $f_n$, the so-called pre-stress term. A detailed derivation of the first two terms can be found in \cite{S-Wyart05}. 

Fully written out in coordinates, the linearised equations of motion in the quasistatic regime (in the absence of damping) are
\begin{align*}
m_i \ddot{\delta \mathbf{r}_i} & = -\sum_j \left[\frac{\partial^2 V_{ij}}{\partial \mathbf{r}_i \partial \mathbf{r}_j} \cdot \delta \mathbf{r}_j + \frac{\partial^2 V_{ij}}{\partial \mathbf{r}_i \partial (R_j \theta_j)} \delta (R_j \theta_j)\right], \\
 \frac{I_i}{R_i^2}\! R_i \ddot{\delta \theta_i} & =\! -\!\sum_j\! \left[\! \frac{\partial^2 V_{ij}}{\partial (\!R_i \theta_i\!) \partial \mathbf{r}_j\!} \!\cdot \delta \mathbf{r}_j\! +\!\frac{\partial^2 V_{ij}}{\partial (\!R_i \theta_i\!) \partial (\!R_j \theta_j\!)} \delta (\!R_j \theta_j\!)\!\right]\!. 
\end{align*}
In the second equation we have used the particle radius $R_i$ to give all coordinates the same dimensions of length. Since the moment of inertia for a cylinder about its central axis is $\frac{1}{2} m_i R_i^2$, the prefactor in the second equation is just $m_i/2$.

Based on these equations, we construct the  dynamical matrix from its $3\times3$ $i,j$ sub-elements between particles $i$, $j$ 
where we are now using the notation $r_{i,\alpha} = (x_i, y_i , R_i \theta_i)$, and similarly $m_{i,\alpha} = (m_i,m_i,m_i/2)$. In simulations \cite{S-henkes2010,S-henkes2016}, we previously set $m_i=1$ for the spatial equations of motion, and even $I_i/R_i^2 = 1$, which corresponds to making the approximation that the particles are all roughly the same size, and are hollow cylinders. 
For this derivation, we will continue to carry the mass and inertia prefactors in order to arrive at a general result that can be used with experimental data. 

We can now derive the $3\times3$ sub-element of the dynamical matrix. Since we have used a local coordinate system to define the local potential, we choose (without loss of generality) $\hat{\mathbf{x}} = \hat{\mathbf{n}}_{ij}$ and $\hat{\mathbf{y}} = \hat{\mathbf{t}}_{ij}$, so that we can write $\delta \mathbf{r} = (\delta x_j -\delta x_i, \delta y_j - \delta y_i$). Later, we will have to rotate this back into a global frame of reference for the full matrix. In this local set of coordinates, the effective potential is 
\begin{align*}
 &V_{ij} = \frac{1}{2} \left[ K_n (\delta x_j - \delta x_i)^2  - \frac{f_n}{r_0}  \left( \delta y_j - \delta y_i \right)^2 +\delta V_{\text{eff}}^f   \right], \\
 & \delta V_{\text{eff}}^f = K_t ((\delta y_j- \delta y_i) - (R_i \delta \theta_i + R_j \delta \theta_j))^2 \quad \text{frictional}, \\
 & \delta V_{\text{eff}}^f = \pm \mu f_n ((\delta y_j- \delta y_i) - (R_i \delta \theta_i + R_j \delta \theta_j)) \quad \text{sliding}.
\end{align*}
As we can readily see, all second derivatives of $\delta V_{\text{eff}}^f$ for sliding contacts are $0$, so that  \emph{sliding contacts do not contribute to the dynamical matrix}. Practically, for sliding contacts, we set $K_t=0$ in the equations below. Then off-diagonal elements of the dynamical matrix are given by 
\begin{equation}
\hat{\mathbf{D}}^{ij} = \frac{1}{\sqrt{m_i m_j}}
 \begin{bmatrix}
  - K_n & 0 & 0 \\
  0 & - K_t + \frac{f_n}{r_0} & K_t/\sqrt{2} \\
  0 & -K_t/\sqrt{2} & K_t/2
 \end{bmatrix}.
 \label{eq:dyn_offdiag}
\end{equation}

The contact $ij$ also contributes to the $ii$ element of the dynamical matrix, as moving particle $i$ itself will also affect its force state. We have a contribution of (note sign changes):
\begin{equation}
\hat{\mathbf{D}}^{ii}_{\text{from} \: j} = \frac{1}{m_i}
 \begin{bmatrix}
   K_n & 0 & 0 \\
  0 &  K_t - \frac{f_n}{r_0} & K_t/\sqrt{2} \\
  0 & K_t/\sqrt{2} & K_t/2
 \end{bmatrix}.
 \label{eq:dyn_diag}
\end{equation}
We then perform rotation of these matrices into a general, global coordinate system, using the angle of the contact normal with the $x$-axis: $\hat{n}_{ij} = (n_x,n_y) = (\cos \phi, \sin \phi)$ and $\hat{t}_{ij} = (n_y,-n_x) = (\sin \phi, - \cos \phi)$.

Since the experimental data is in a form where each contact is single-counted only, this means we also need to use the information above to construct the $ji$ contact, and add to the $jj$ diagonal element. The derivation for these is identical, except for swapping the $j$ and $i$ labels, so in their local coordinates the results of Eq.~\ref{eq:dyn_diag} and Eq.~\ref{eq:dyn_offdiag} are identical. 

However, more subtly, $\hat{\mathbf{n}}_{ji}=-\hat{\mathbf{n}}_{ij}$ and $\hat{\mathbf{t}}_{ji}=-\hat{\mathbf{t}}_{ij}$, while the axis of rotation in our coordinate system ($z$) is not affected. Since the $nn$, $nt$, $tn$ and $tt$ terms have two unit vector components, the signs cancel and nothing changes. For the $\alpha\alpha$ term, nothing changes. However, for the off-diagonal components $n\alpha$, $\alpha n$, $t\alpha$ and $\alpha t$, the sign \emph{will} change. 
Then \emph{if}  we use the $\hat{\mathbf{n}}_{ij}$ and $\hat{\mathbf{t}}_{ij}$ as a basis for the flipped contacts as well, we have the following contributions:
\begin{equation}
\hat{\mathbf{D}}^{ij} = \frac{1}{\sqrt{m_i m_j}}
 \begin{bmatrix}
  - K_n & 0 & 0 \\
  0 & - K_t + \frac{f_n}{r_0} & -K_t/\sqrt{2} \\
  0 & K_t/\sqrt{2} & K_t/2
 \end{bmatrix}
 \label{eq:dyn_offdiag_flip}
\end{equation}
The contact $ji$ also contributes to the self-element $jj$ of the dynamical matrix and we have a contribution
\begin{equation}
\hat{\mathbf{D}}^{jj}_{\text{from} \: i} = \frac{1}{m_j}
 \begin{bmatrix}
   K_n & 0 & 0 \\
  0 &  K_t - \frac{f_n}{r_0} & -K_t/\sqrt{2} \\
  0 & -K_t/\sqrt{2} & K_t/2
 \end{bmatrix}
 \label{eq:dyn_diag_flip}
\end{equation}

The dynamical matrix for frictional packings is implemented in the \emph{Hessian} class of the rigid analysis python library \cite{S-rigid}.

\subsection{Mapping out rigid regions}

\begin{figure}
	\centering
	\includegraphics[width=0.48\textwidth]{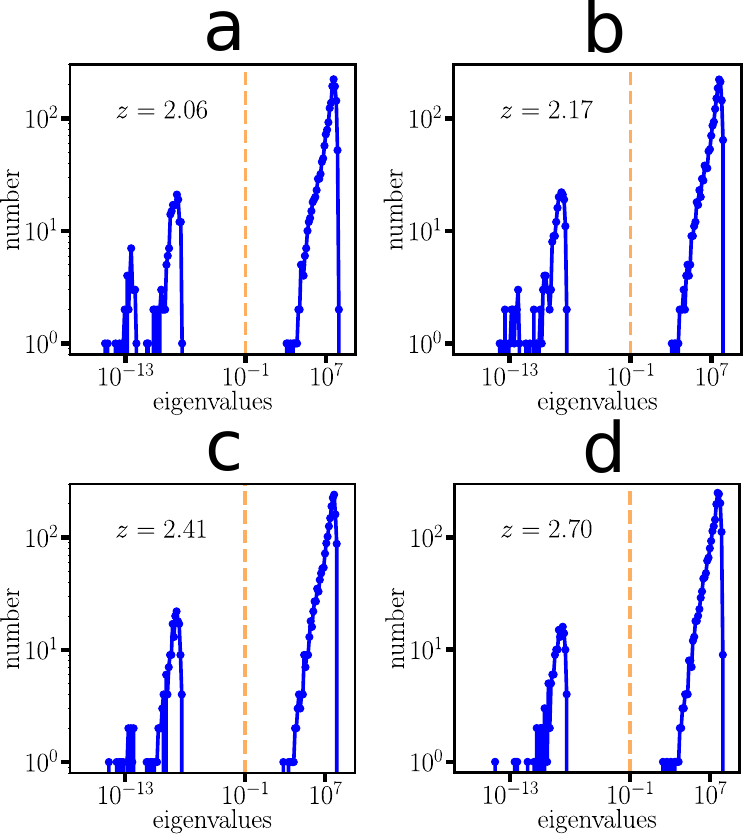}
	\caption{Distribution of eigenvalues for four different samples with average coordination number shown in figures.}
	\label{fig:eigval_SI}
\end{figure}

Now that we have constructed the dynamical matrix, we can compute all its eigenvalues and eigenvectors accordingly. In Fig.~\ref{fig:eigval_SI}, we show distributions of computed eigenvalues for four samples with a range of $z$. From each plot we observe two peaks at approximately $10^{-8}$ and $10^7$ and a wide gap between them. We, therefore, treat eigenvalues below $10^{-1}$ as zeros with their corresponding eigenvectors representing zero modes labeled as $(\mathbf{r}_{i,k},\theta_{i,k})$ where $i$ labels different particles and $k$ labels different zero modes. From all $N_{\text{zero}}$ zero modes we compute the relative translational and rotational displacements between particles $i$, $j$ using

\begin{align}
\delta r_{i,j,\text{trans}}^2 &= \frac{1}{N_{\text{zero}}} \sum_k \left[ ((\mathbf{r}_{j,k}\!-\!\mathbf{r}_{i,k}) \!\cdot\! \hat{\mathbf{n}})^2 +   ((\mathbf{r}_{j,k}\!-\!\mathbf{r}_{i,k}) \!\cdot \! \hat{\mathbf{t}} \! )^2 \right] \\
 \delta \theta_{ij,\text{rot}}^2 &=  \frac{1}{N_{\text{zero}}}  \sum_k \left(\theta_{i,k}+\theta_{j,k} \right)^2.
\end{align}

\begin{figure}
	\centering
	\includegraphics[width=0.4\textwidth]{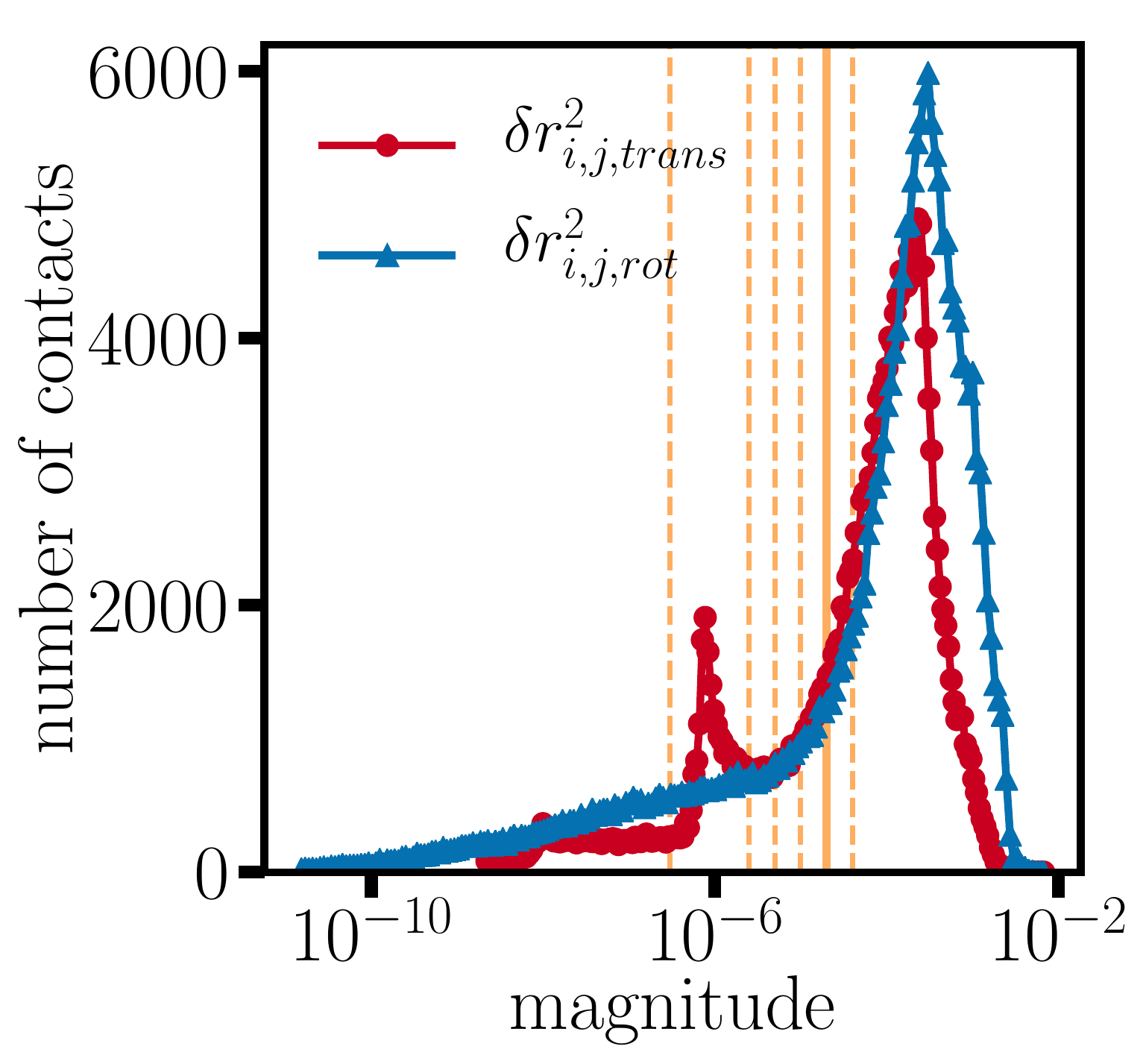}
	\caption{A semi-log plot of the distribution of all relative translational displacements (red) and rotational displacements (blue) between all pairs of particles for all samples, as a function of displacement magnitude, increasing exponentially from left to right. From left to right,the six orange lines label six rigidity thresholds, $3\times 10^{-7}$, $2.5\times 10^{-6}$, $5\times 10^{-6}$, $1\times 10^{-5}$, $2\times 10^{-5}$,  $4\times 10^{-5}$. Contacts on the right side are treated as floppy and contacts on the left side are rigid. The solid line is the threshold used in main text.}
	\label{fig:motion_distribution_SI}
\end{figure}

\begin{figure}
 \centering
 \includegraphics[width=0.48\textwidth]{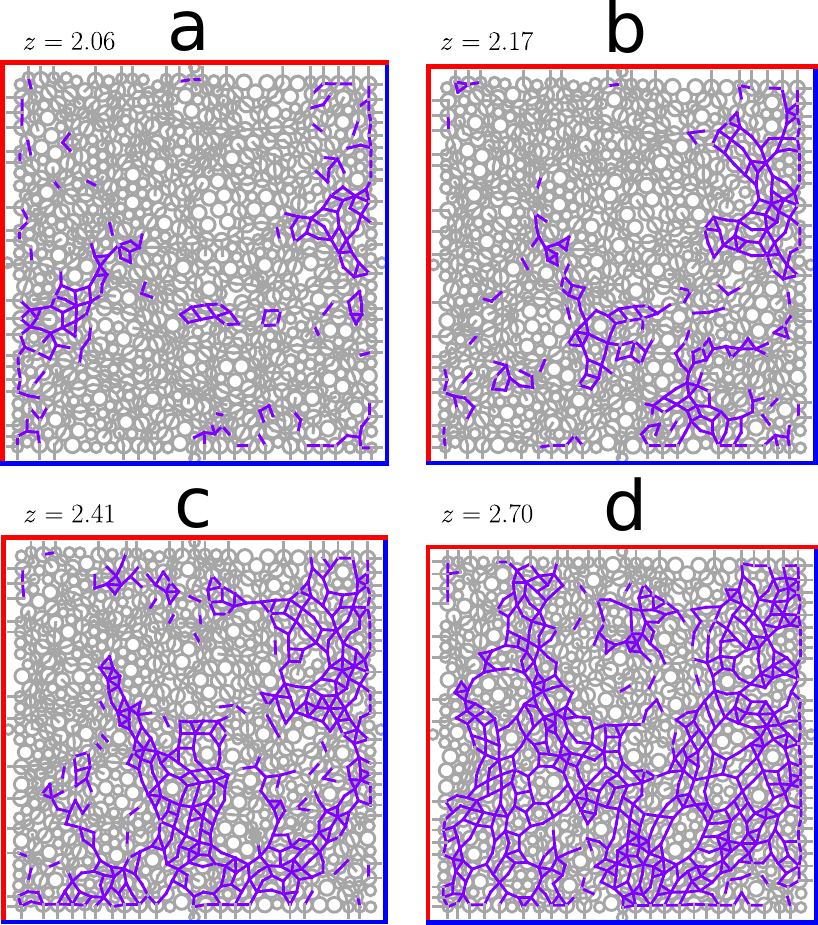}
 \caption{Representative identification of rigid regions for the same four samples in Fig.~\ref{fig:eigval_SI}, using the dynamical matrix and rigidity threshold set to $2\times 10^{-5}$. Purple contacts are in rigid regions under both translational and rotational considerations;  grey contacts are floppy. }
 \label{fig:rigid_region_SI}
\end{figure}

In Fig.~\ref{fig:motion_distribution_SI}, we plot distributions of $\delta r_{i,j,\text{trans}}^2$ and $ \delta \theta_{ij,\text{rot}}^2 $ respectively for all contacts and observe two peaks in relative translational displacements.  Based on this, we select an appropriate rigidity threshold $\tau$ such that any relative motion below that value is treated as zero. These are considered rigid and form the rigid regions shown in Fig.~\ref{fig:rigid_region_SI}. The sensitivity of our results to this threshold choice is discussed in an upcoming section. 

\clearpage
\section{Determination of parameters \label{sec:params}}

\subsection{Particle parameters}
To compute particle masses, we note that the Vishay particles are disk-shaped and have a density of $\rho = 1.06\,\mathrm{g/cm}^3$. For particle radii corresponding to $R_1 = 5.5$~mm and  $R_2 = 7.6$~mm, and for a thickness $h = 3.1$~mm, the particle masses are then
$m_1 = 3.12 \times 10^{-4} \mathrm{kg}$, and $m_2 = 5.96 \times 10^{-4} \mathrm{kg}$. 
The raw data from images are measured in pixels, with radii $R_1 = 20$ and $R_2 = 29$~pixels, and we can therefore deduce a global conversion factor of $c = 2.7 \times 10^{-4}$~m/pixels. 

To estimate the stiffness coefficients, we read off the dynamical matrix equations that their units are acceleration per length, i.e. $[K_n]= [M][T^{-2}]$. 
Using Hertzian contact theory~\cite{S-johnson_contact_1985}, for two cylinders with parallel axes,
the force is given approximatively by $F = \frac{\pi}{4} E^* h d$, where $d$ is the indentation depth (overlap), and $E^*$ is the scaled Young's modulus, 
\begin{equation} \frac{1}{E^*} = \frac{1-\nu_1^2}{E_1} + \frac{1-\nu_2^2}{E_2},\end{equation} 
where $E_i$ and $\nu_i$ are Young's moduli and the Poission ratios of the two materials, respectively. If we assume that $\nu_1=\nu_2 = 0.5$ (i.e., an incompressible material), and set $E_1 = E_2 = E$, we find $E^* = \frac{2}{3} E$. Finally,
 we obtain $K_n = \frac{\pi}{6} E h$.  For an order of magnitude estimate, the
Young's modulus of Vishay is $E \approx 4\,\mathrm{MPa}$, and since $[E] = [M][T^{-2}][L^{-1}]$, our units are correct.
For interactions between two particles, the stiffness coefficient of the harmonic elastic interaction related to normal forces is therefore $K_n=6490 ~\mathrm{kg \cdot s}^{-2}$. For simplicity, we will assume that we have a Cundall-Strack-like
relation for tangential motion, so that $K_t = K_n$.

The boundary of the system is significantly larger and heavier than a single particle, so that both the mass $m_b$ and the moment of inertia $I_b$ are much larger. Using a rough order of magnitude estimate of boundary size and shape, we arrive at $m_b \approx 10^2 (m_1+m_2)/2$ and $I_b \approx 10^4 (I_1+I_2)/2$. Therefore, the terms in the dynamical matrix corresponding to the particle-boundary and boundary-boundary interactions will be very small, and the corresponding displacements in the normal modes are also tiny. Compared to a system without boundaries, this softly enforces the constraint of an immovable boundary. 
For contacts between particles $p$ and a wall $w$ made of dissimilar materials with Young's moduli $E_p$ and $E_b$ and Poisson ratio's $\nu_p$ and $\nu_w$, the effect modulus that enters the stiffness coefficient or a particle-wall contact is  $\frac{1}{E^*_{pw}} = \frac{1-\nu_p^2}{E_p} + \frac{1-\nu_w^2}{E_w}$. For wall stiffness $E_w$ lying between $E_p$ and infinity, we obtain  $E^*_{pw} = 1.0 E^*_{pp}-2.0 E^*_{pp}$, in practice the wall is significantly stiffer than the particles.   In the dynamical matrix, we use the same $K_n$ and $K_t$ for particle-boundary contacts as for particle-particle contacts.

\begin{figure}[h]
	\centering
	\includegraphics[width=0.4\textwidth]{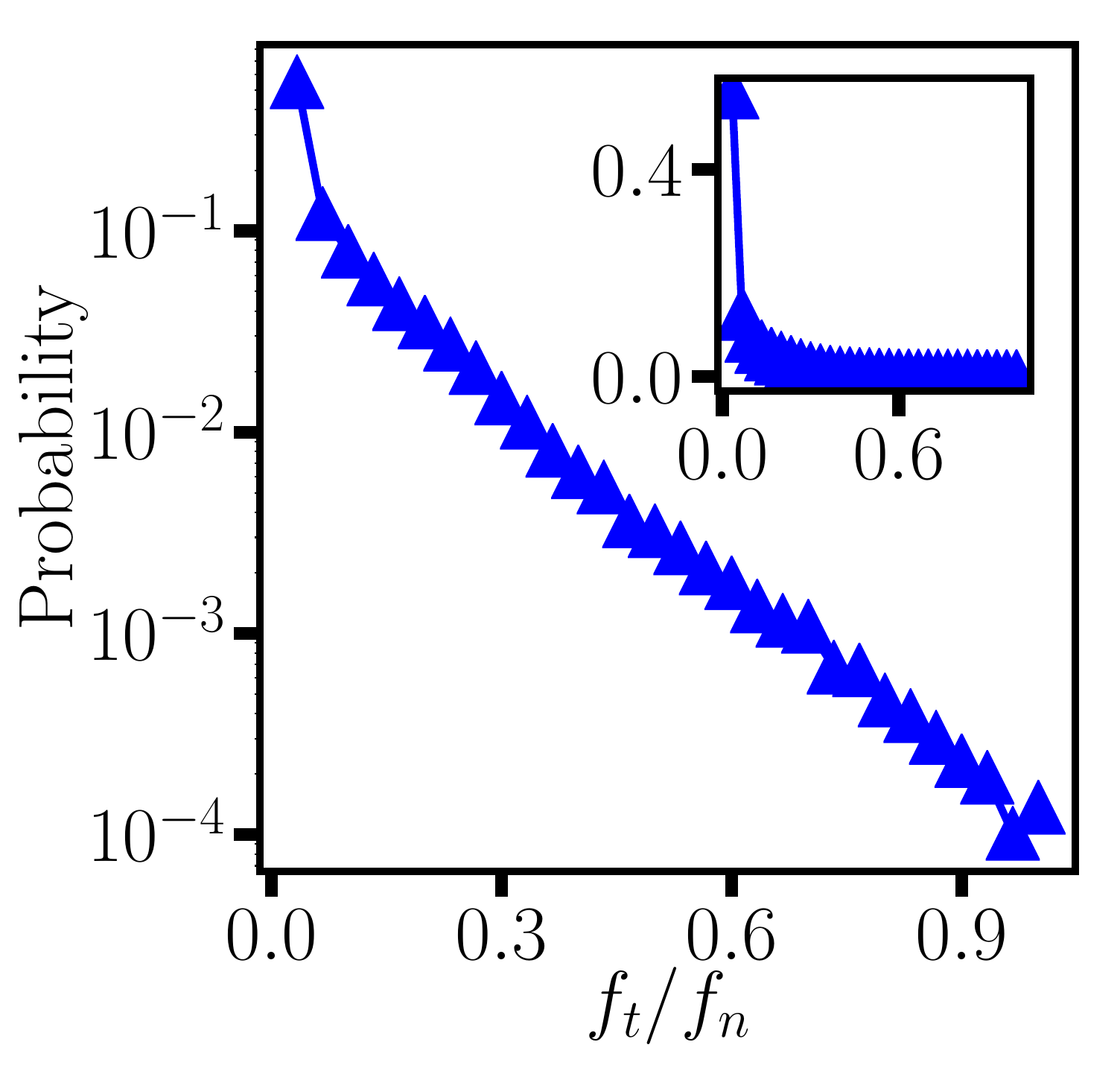}
	\caption{Semi-log plot of the probability distribution of $f_t/f_n$ from all packings. Inset: same, on a linear axis scale.}
	\label{fig:ftfnratio}
\end{figure}

In both methods, the pebble game and the dynamical matrix, identifying whether a contact is sliding is essential and we therefore need an estimate for the friction coefficient $\mu$. In Fig.~\ref{fig:ftfnratio} we plot the probability distribution of the ratio of tangential forces to the corresponding normal forces, or \emph{mobilisation}. While we do not see an accumulation of probability near the Coulomb threshold as in some simulated packings \cite{S-shundyak2007force}, this figure suggests that the friction coefficient is approximately $\mu=0.3$ since the probability drops below $10^{-2}$ around a ratio of $0.3$. With these parameters, on average $5.2\%$ of all contacts are sliding contacts. Below, we test how our numerical methods are affected by the choice of friction coefficient.

\subsection{Pressure calculation}
The pressure on a given particle $i$ due to its contact forces with particles $j$ is derived from the virial part of the Irving-Kirkwood stress tensor \cite{S-irving1950statistical},
\begin{equation} \hat{\mathbf{\sigma}}_i = \frac{1}{A_i} \sum_{j} \mathbf{r}_{ij} \otimes \mathbf{F}_{ij}, \end{equation}
where $A_i$ is the area (in two dimensions) of the plane associated to particle $i$ in a tesselation so that $\sum_i A_i = A$, the total system size. This is to ensure that the stress is an intensive quantity.
With $f^n_{ij}$ denoting the normal force between particle $i$ and $j$, the local pressure on particle $i$, i.e. the trace of the stress tensor,  is then given $p_i=\frac{1}{A_i} \sum_{j} f^n_{ij} r_{ij}$, where $A_i$ is the area possessed by particle $i$ after Voronoi tessellation. When considering every particle $i$ in rigid cluster and its neighbor $j$, we then obtain the pressure within the rigid cluster as $p_{\text{in}}=\frac{\sum_{i} \sum_j f^n_{ij} r_{ij}}{\sum_{i} A_i}$. Note that a particle is in the rigid cluster if any one bond connected to it is identified as a rigid bond. Similarly, we can compute the pressure outside the rigid cluster.

\subsection{Sensitivity to parameter choice: $\mu$, $\tau$}

Since the decomposition of rigid clusters and the construction of dynamical matrix depend on the choice of both the friction coefficient $\mu$ and the rigidity threshold $\tau$ (for the dynamical matrix only), we need to test to what extent our results depend on those parameters. Here, we put our choice of $\mu=0.3$ and $\tau=2\times10^{-5}$ (corresponding to Fig.2a-c in the main paper) into context. We vary $\tau$ from $3\times 10^{-7}$ to $4\times 10^{-5}$, corresponding to the cutoffs in the displacement magnitude indicated in Fig. \ref{fig:motion_distribution_SI}. We also test the dependence on the friction coefficient by performing the analysis with $\mu=0.2$ instead of $\mu=0.3$. 

\begin{figure*}
	\centering
	\includegraphics[width=1.0\textwidth]{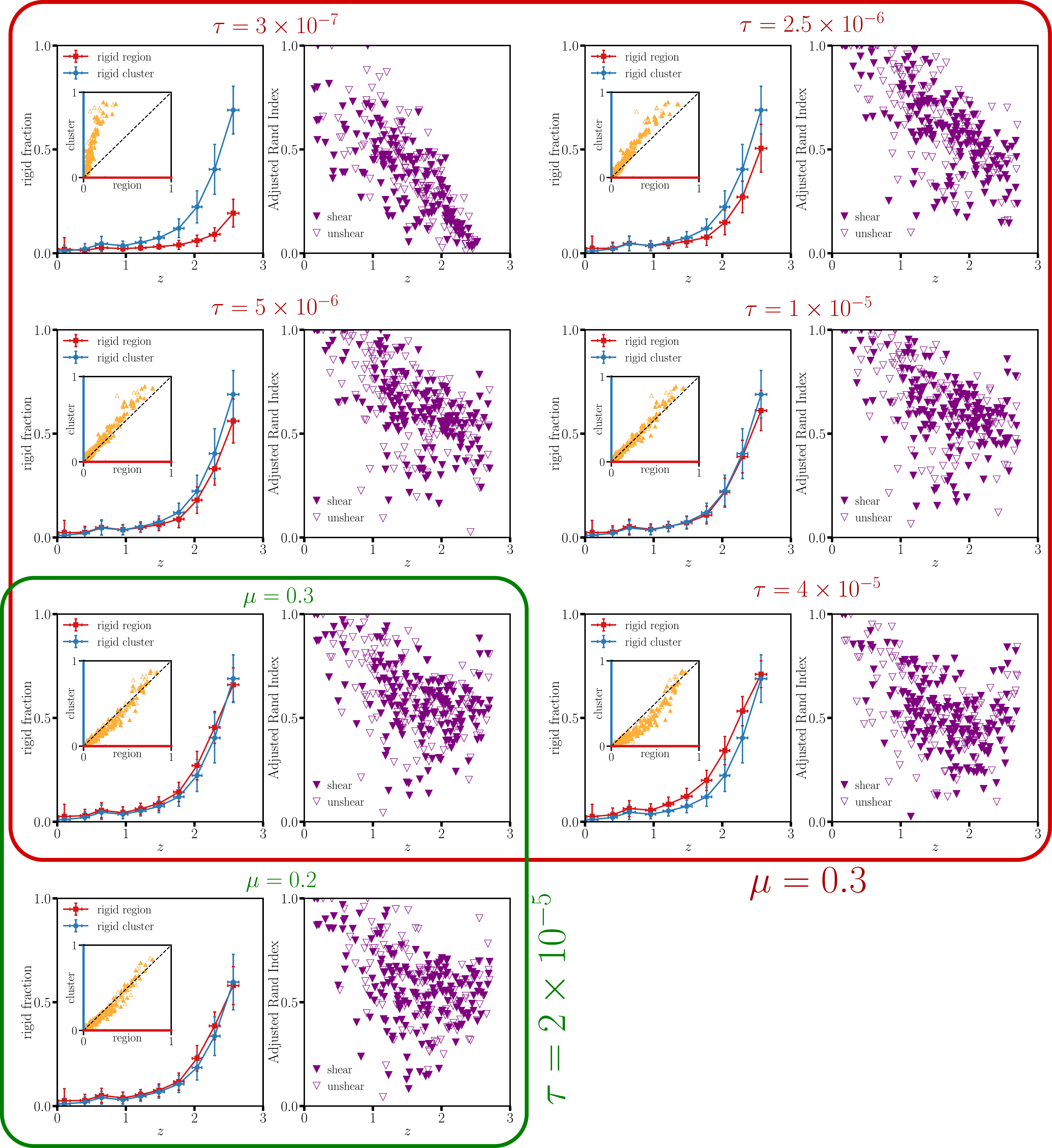}
	\caption{Robustness of the rigid cluster decomposition and rigid region analysis to changes in displacement threshold $\tau$ and friction coefficient $\mu$. For each set of thresholds, we compute the equivalent of Fig.2a-c in the main text, i.e. the rigid fraction as a function of $z$, and the correlations between rigid regions (here shown as insets) and rigid clusters using the adjusted Rand index. Outlined in red: effect of changing the displacement threshold from $\tau = 3\times 10^{-7}$ to $\tau = 4\times 10^{-5}$. Outlined in green: effect of changing the friction coefficient from $\mu = 0.3$ to $\mu=0.2$. The values shown in Fig.2a-c of the main text are at the intersection of the red and green sets.}
	\label{fig:fraction_mu_tau_vary}
\end{figure*}

In Fig.~\ref{fig:fraction_mu_tau_vary}, we plot the same analysis as that presented in Fig.2a-c in the main text for each set of parameters. At the extreme end, for $\tau=3\times 10^{-7}$ i.e. a threshold to the left of both peaks in Fig.~\ref{fig:motion_distribution_SI}, we see that the fraction of rigid region decreases substantially due to the rigidity threshold being too small and the correlation between two methods at higher $z$ disappears. However, above this threshold in $\tau$, we observe that the correlation between the rigid clusters and the rigid regions is robust to changes in  $\tau$ parameter space corresponding to the ``valley'' in relative translational displacement distribution shown in Fig.  \ref{fig:motion_distribution_SI}.

For the lower friction coefficient $\mu=0.2$, both the rigid region fraction and rigid cluster fraction are decreased compared to $\mu=0.3$, which can be explained by the fact that the Coulomb threshold is lower, leading to more fully mobilized sliding contacts and so more motion is allowed in both methods. The correlations between both measures remains robust however, as can be seen in the graph of the Adjusted Rand Index (see \S\ref{sec:rand}).

\section{Adjusted rand index (ARI) \label{sec:rand}}

The Rand index \cite{S-Rand1971} is a commonly-used statistical measure to quantify the degree of similarity between two different data clusterings. For a given set of $n$ elements $E=\{e_1,e_2,e_3,...,e_n\}$, two clustering methods obtain two partitions of $E$, call them $X=\{x_1,x_2,...,x_r\}$ and $Y=\{y_1,y_2,...,y_s\}$. Every element in $X$ and $Y$ is a subset of $E$. For each pair of elements in $E$, there are four cases:
\begin{enumerate}[a]
	\item in same subset of $X$ and in the same subset of $Y$
	\item in same subset of $X$ but in different subsets of $Y$
	\item in different subsets of $X$ but in the same subset of $Y$
	\item in different subsets in $X$ and in different subsets of $Y$
\end{enumerate}
The cases $a$, $b$, $c$, $d$ together count the total number of pairs of elements,  $\frac{n(n-1)}{2}$.  The Rand index is then defined as the fraction 
\begin{equation}
\mathrm{RI} =\frac{a+d}{a+b+c+d}.
\end{equation}
By definition, RI is a number between $0$ and $1$, where $0$ signifies maximum anti-correlation and $1$ signifies maximum correlation. For random clustering, $\mathrm{RI} = 0.5$. 

Since the Rand Index is computed by counting permutations, once the number of clusters or the size distribution of those clusters vary drastically, for example in low $z$ cases and high $z$ cases in our project, RI cannot capture the correlation between two clustering methods effectively. We therefore use the Adjusted Rand Index (ARI) \cite{S-Hubert1985} to remove such effects, in which the cases (a,b,c,d) are tabulated the same way but the number of possible combinations are taken into account:
\begin{align}
\mathrm{ARI} =\frac{{n \choose 2}(a+d)-(a+b)(a+c)-(d+b)(d+c)}{{n \choose 2}{n \choose 2}-(a+b)(a+c)-(d+b)(d+c)}
\end{align}
Unlike the RI, the ARI takes values from $-1$ to $1$, with $0$ corresponding to random clustering. 

In Fig.~\ref{fig:fraction_mu_tau_vary} we also plot ARI for each set of parameters discussed in the last section in \S\ref{sec:params}. Except for the extreme end of $\tau=3\times 10^{-7}$, these ARI plots show robust correlation between the two rigid analysis methods.

\clearpage

\end{document}